\documentclass[aps,prb,twocolumn,showpacs]{revtex4}

\usepackage{graphicx}
\DeclareGraphicsExtensions{.png,.jpg,.eps}

\usepackage{xcolor}
\usepackage{amsmath}

\newcommand{\STMEPR}{Baumann2015,natterer2017,choi2017,yang2017,willke2018,willke2018b,bae2018,yang2018,yang2019,willke2019}

\begin{document}

\title{Single spin   resonance driven by electric modulation of the $g$ factor anisotropy}

\author{ A. Ferr\'on$^{1}$,  S. A. Rodr\'iguez$^{1}$, S. S.  G\'omez$^{1}$, J. L. Lado$^{2}$,   J. Fern\'andez-Rossier$^{3}$
\footnote{On leave from Departamento de Fisica Aplicada, Universidad de Alicante, 03690 Spain }
}
\affiliation{ 
(1) Instituto de Modelado e Innovaci\'on Tecnol\'ogica (CONICET-UNNE) and 
Facultad de Ciencias Exactas, Naturales y Agrimensura, Universidad Nacional 
del Nordeste, Avenida Libertad 5400, W3404AAS Corrientes, Argentina.
\\(2) Department of Applied Physics, Aalto University, Espoo, Finland. 
\\(3) QuantaLab, International Iberian Nanotechnology Laboratory (INL),
Av. Mestre Jos\'e Veiga, 4715-330 Braga, Portugal. 
}

\date{\today}

\begin{abstract}
We address the problem of  electronic and nuclear  spin resonance  of an individual atom on a surface driven by 
a scanning tunnelling microscope.    Several mechanisms have been proposed so far, some of them based on the modulation of exchange and crystal field  associated to a piezoelectric displacement of the adatom driven by the RF tip electric field.
	Here we consider a new mechanism,  where the piezoelectric displacement modulates   the $g$ factor anisotropy, leading both to electronic and nuclear spin flip transitions.  We discuss thoroughly the cases of  Ti-H ($S=1/2$)  and Fe $(S=2)$  on MgO, relevant for recent experiments.  We model the system using two approaches. First, an analytical model that includes  crystal field,  spin orbit coupling and hyperfine interactions.    Second, we carry out density functional based calculations. We find that 
 the modulation of the anisotropy of the $g$ tensor due to the piezoelectric displacement of the atom is an additional mechanism for STM based single spin resonance, that would be effective in $S=1/2$ adatoms with large spin orbit coupling.   In the case of Ti-H on MgO, we predict a modulation  spin resonance frequency driven by the DC electric field of the tip.

\end{abstract}

\maketitle

\section{Introduction}
The quest of single spin electron paramagnetic resonance (EPR) driven with a 
scanning tunneling microscope (STM) has been pursued for many years
\cite{Manasen_Hamers_prl_1989,reviewbalatsky2012}. The first report of STM-ESR
of individual adatoms on a surface of MgO(100)/Ag   \cite{Baumann2015} has 
been followed by  several dramatic breakthroughs in the study of spin physics 
of individual magnetic atoms\cite{natterer2017,choi2017,yang2017,willke2018,
willke2018b,bae2018,yang2018,yang2019,willke2019,willke2019b}.  
This technique permits to 
carry out absolute measurements of the magnetic moment of individual 
atoms\cite{natterer2017,choi2017}. 
The spectral resolution achieved so far, down to a few MHz, has made it 
possible to resolve the  hyperfine structure of Fe, Ti and Cu 
atoms\cite{willke2018b,yang2018}.  In the case of Cu adatoms, the electrical 
driving of nuclear spin-flip transitions that preserve the electronic spin has 
been demonstrated as well\cite{yang2018}. Thus, STM-EPR permits to drive  
the electronic and nuclear spins of individual atoms on surfaces, as well as 
artificially created structures, such as dimers\cite{yang2017,bae2018}.  
Importantly, the STM-ESR technique is being now implemented in several different  
laboratories, at higher  temperatures\cite{natterer2019} and higher driving 
frequencies \cite{seifert2019}.

An important question  in the STM-EPR context\cite{{\STMEPR}}, and also for 
experiments  reporting electric control of individual nuclear spin in  single 
molecule  transport \cite{thiele2014,godfrin2017}, is the understanding of  
how  electric fields couple both to  electronic and nuclear  spin degrees of 
freedom. This question has also been addressed in other systems. The idea of 
electric dipole spin resonance was proposed back in 1960 by 
Rashba\cite{rashba1960}.
 Electrical control of spin qubits has been reported  in semiconductor nanostructures, based both 
 on modulation of the $g$ factor\cite{kato2003} and on inhomogeneous magnetic 
 fields\cite{Tokura06,pioro2008}.  Electric fields have been used to drive 
 spin resonance of itinerant electrons in InSb\cite{bell62} and localized 
 magnetic dopants in ZnO\cite{george13}.

In the seminal paper of Baumann {\em et al}\cite{Baumann2015}, where the first 
STM-EPR experiment was carried out with Fe atoms on an  MgO surface,  a 
mechanism was proposed to account for the coupling of the STM voltage to 
the {\em electronic} spin, that depended on the specific details of the 
microscopic Hamiltonian of that system.  The mechanism is based on the 
assumption that the $RF$ field induces a vertical piezoelectric displacement 
of the adatom, $\delta z\propto eV_{RF}$, that in turns modifies the crystal 
field Hamiltonian of the $d$ orbitals of Fe.  This modulation, together with 
spin-orbit coupling and a strong in-plane Zeeman field, would lead to spin 
transitions between the two lowest energy states of the $S=2, S_z=\pm 2$ 
of Fe, a non-Kramers doublet integer spin system \cite{hoffman1994}.
 
Other  mechanisms have been proposed to account for the driving of the surface 
spin by the tip bias voltage\cite{reviewbalatsky2012,berggren2016,lado2017,
shakirov2019,galvez2019}. For instance, in Ref. [\onlinecite{lado2017}] we 
proposed a  mechanism  based on the modulation of the exchange interaction 
between the magnetic tip and the magnetic adatom,  that originates also from 
the  piezoelectric distortion of the adatom.

Here we propose another complementary mechanism, that can coexist with the 
others, based on the electric modulation of the $g$ tensor associated to the 
piezoelectric distortion of the adatom. As in the case of the crystal 
field\cite{Baumann2015} and exchange\cite{lado2017,yang2019} mechanisms, we 
also assume that the magnetic adatom undergoes a piezoelectric displacement.  
In turn, this modulation changes the crystal field parameters that control 
the  anisotropy of the  electronic spin interactions, that leads to 
an anisotropic $g$ factor and to a renormalization of the hyperfine coupling.
As we show below, these  modulations lead both to electronic and nuclear spin 
flip transitions. 

The rest of this paper is organized as follows. 
In section \ref{general}  we present a general argument to show that an anisotropic  time dependent modulation of the $g$ tensor of a $S=1/2$ system leads to electronic spin transitions.
In section \ref{sec:c4model} we briefly present a single particle Hamiltonian 
for  a $d^1$ adatom with $C_4$ symmetry, valid for Ti-H adatom on the oxygen 
site of MgO(001). 
In section \ref{sec:DFT} we present our description of the Ti-H adatom on MgO 
based on Density functional theory (DFT) calculations and how this connects 
with the crystal field Hamiltonian presented in the previous section.

In section \ref{sec:gfacmodel} we derive analytical expressions for the $g$ 
tensor anisotopy of Ti-H on MgO,  based on the model of section 
\ref{sec:c4model}.  The $g$ tensor obtained depends on the Ti  spin-orbit 
coupling and the crystal field parameters, that can be obtained from DFT. In 
section \ref{sec:piezog} we discuss how the $g$ factor can be modulated for 
Ti-H on MgO  by application of an electric field between tip and surface and 
we compute the associated Rabi energy. In section \ref{sec:Fe} we briefly 
present the analogous piezoelectric modulation for Fe on MgO. In section 
\ref{sec:hyper} we discuss how the contact hyperfine interaction become 
anisotropic due to the $g$ factor anisotropy and how the $g$ factor modulation 
could induce nuclear spin flip transitions. In section \ref{sec:exch}  we 
discuss the role of both the $g$ tensor anisotropy of the adatom and the 
magnetic anisotropy of the tip in the efficiency of the exchange modulation
\cite{lado2017} mechanism. In section \ref{sec:shift} we show that the DC 
component of the tip-surface electric field induces a shift of the transition 
energy of the adatom due to the modification of the $g$ tensor. Finally, in 
section \ref{wrap} we present some limitations of our models and we list our 
main conclusions. The appendices describe technical steps  of some results 
used in the main text. 

\section{ Spin transitions driven by anisotropic modulation of the $g$ tensor\label{general}}
 For a  free electron in vacuum, the interaction with a magnetic field is 
perfectly isotropic, in the sense that the energy splitting is the same 
regardless of the direction of the magnetic field $\vec{B}$. This results 
leads to the  isotropic Zeeman interaction, $g\mu_B \vec{S}\cdot\vec{B}$ . In 
contrast, for a general class of systems,   the interplay between the spin 
orbit coupling $\vec{\ell}\cdot\vec{S}$,  the orbital  coupling to the magnetic field $\vec{\ell}\cdot\vec{B}$,  and the crystal field splitting leads to  an anisotropic Zeeman 
interaction. For instance, in the case of $S=1/2$  adatoms, such as Ti-H 
\cite{yang2017,willke2018b,bae2018,yang2018,yang2019} and Cu\cite{yang2018} on 
a $001$  MgO surface, the interplay between the spin orbit coupling and the 
crystal field splitting leads to an anisotropic Zeeman interaction with 
different off-plane ($z$) and in-plane $xy$\footnote{Given the $C_4$ symmetry 
of the adatom on the oxygen position in MgO, $x$ and $y$ directions are 
equivalent and we assume $B_y=0$}:
\begin{equation}
{\cal H}_Z= g_x \mu_B B_x S_x  + g_z \mu_B B_z S_z =\mu_B \vec{b}_0\cdot\vec{S}
\label{h00}
\end{equation}
\noindent where $\vec{b}_0=(g_xB_x,0,g_zB_z)$.

As we show below, the tip $ac$ electric field modulates the $g_x$ and $g_z$ 
coefficients, resulting in a time dependent perturbation: 
\begin{equation}
{\cal V}(t) = (\delta g_x  \mu_B B_x S_x  + \delta g_z \mu_B B_z S_z)\cos (2\pi f t)
\label{V}
\end{equation}
This equation can be written down as 
\begin{equation}
{\cal V}(t)=\cos (2\pi f t)\mu_B \vec{b}_1\cdot\vec{S}
\label{V2}
\end{equation}

\noindent where $\vec{b}_1=(\delta g_xB_x,0,\delta g_zB_z)$. This 
perturbation can induce  spin transitions between the two eigenstates of 
${\cal H}_0$ if $\vec{b}_0$ and $\vec{b}_1$ are non-collinear, 
$|\vec{b}_1\times \vec{b}_0|\neq 0$. This  yields 
\begin{equation}
 \frac{\delta g_z}{g_z} \neq \frac{ \delta g_x}{g_x}
 \label{mc}
 \end{equation}
Thus, the perturbation (\ref{V}) induces spin transitions if the relative 
modulations of the $g$ factor are {\em different}. If we express the 
perturbation Hamiltonian in the basis of  eigenstates of 
${\cal H}_Z|\pm\rangle=\pm\frac{\Delta_Z}{2}|\pm\rangle$:
\begin{equation}
{\cal V}(t)=\Omega_g \cos (2\pi f t)\left( |+\rangle\langle-|+  |-\rangle\langle+|\right)
\label{Rabi0}
\end{equation}
where the Rabi coupling $\Omega_g$ is given by particularly simple equation,
derived in  appendix \ref{Rabiapp}: 
\begin{equation}
\Omega_g=\frac{\Delta_Z}{4}  \sin 2\theta
 \left( \frac{\delta g_z}{g_z}   - \frac{\delta g_x}{g_x}   \right)
\label{Rabi1}
\end{equation}
where $\Delta_z\equiv \mu_B |\vec{b}_0|$ is the Zeeman splitting, $\theta$ is 
the polar coordinate of the $\vec{b}_0$ defined in Eq. (\ref{h00}) (see 
also Eq. (\ref{b0}), Eq. (\ref{sin}) and Eq.(\ref{cos})). From Eq. (\ref{Rabi1}) we 
immediately infer that the Rabi coupling created by the modulation of the $g$ 
factor  scales linearly with the {\em magnitude} of the magnetic field, and 
has a very strong dependence on its orientation relative to the normal of the 
surface.

\section{ A model  Hamiltonan for T${\rm i }$-H on M${\rm g }$O
\label{sec:c4model}}
We now consider a toy model  that describes a single electron occupying a $d$ 
shell with a crystal field splitting with   $C_4$ symmetry for rotations in 
the $xy$ plane around the $z$ axis.   This permits to obtain closed analytical 
expressions for the $g$ tensor in terms of the crystal field parameters and 
the spin orbit coupling. In addition,  our DFT calculations, discussed below, 
show that the model provides a fairly good description of hydrogenated Ti 
adatoms on the oxygen site of an MgO surface, relevant for STM-EPR 
experiments\cite{yang2017,willke2018b,yang2018,bae2018,yang2019}.

\begin{figure}[hbt]
\includegraphics[width=1.1\linewidth]{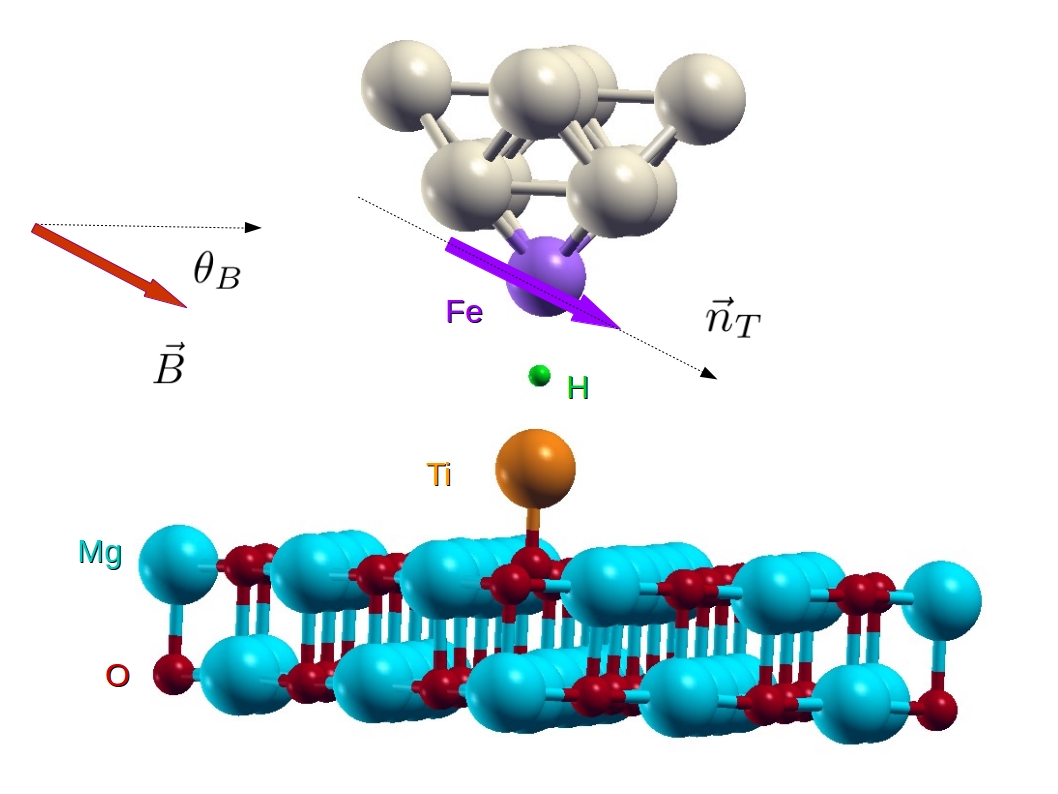}
\caption{\label{fig1}   Scheme of  the STM-ESR experimental set-up
\cite{\STMEPR} with the atomic structure of Ti-H 
on the oxygen site of MgO. $\vec{B}$ is the external magnetic field applied 
during the experiment (with an angle $\theta_B$ measured with respect to the 
surface) and $\vec{n}_T$ shows the direction of the magnetic moment of the tip. 
Red balls represent O atoms, blue balls Mg atoms, violet ball is for Fe atom, 
green ball for H atom and the orange one is for Ti atom.}
\end{figure}

Ti$^{2+}$ on MgO has 2 electrons in the $d$ shell, and our DFT calculations 
show it has $S=1$, in contrast with the experimental results
\cite{yang2017,willke2018b,bae2018,yang2018,yang2019}.
It has been proposed  that the reason why Ti/MgO has $S=1/2$ is because it  
chemisorbs an hydrogen atom\cite{yang2017}. Our DFT calculations  back up this 
assumption\cite{yang2017}. They show that hydrogen sits on top of Ti, almost co-linear with 
the Oxygen-Ti line that goes perpendicular to the surface.  In that geometry, 
the $s$ orbital of $H$ hybridizes  both with $d_{z^2}$  and $s$ orbital of Ti and forms a molecular 
bonding-anti-bonding pair that hosts 2 electrons.  This leaves only a single electron in the $d$ shell, 
that occupies the $x^2-y^2$ orbital, so that the  Ti-H system effectively has $S=1/2$.
We use the following Hamiltonian for  the outermost electron of a single 
$S=1/2$ electron in a $d$ shell, that includes crystal field terms, spin orbit 
coupling and Zeeman interaction:
 \begin{eqnarray}
 {\cal H}_0= -|D|\ell_z^2 
	 + F\left((\ell^{(+)})^4+(\ell^{(-)})^4\right) +\nonumber \\
 +\lambda \vec{S}\cdot\vec{\ell}+ \mu_B \vec{B}\cdot\left( g\vec{S}+\vec{\ell}\right)
 \label{H}
 \end{eqnarray}
Here, $\ell$ is the single particle  angular momentum operator for the $d$ electrons and $\vec{S}$ 
are the spin 1/2 matrices.  Notice that Baumann {\em et al.}\cite{Baumann2015} 
used a mathematically similar expression for a multi-electronic Hamiltonian 
multiplet with $L=S=2$, valid for Fe on MgO.  
The crystal field terms account for the electrostatic  interactions of 
the first neighbour charged ions of the Ti adatom (see Fig. \ref{fig1}). 
Mg atoms  are positively charged ions that reduce the energy of the $xy$ 
and $x^2-y^2$ orbitals compared to the $xz$, $yz$ and z$^2$ orbitals. Oxygen
atom is negatively charged and it increase the energy of the $z^2$ orbital. 
The $D$ term accounts for these effects. In addition, the $F$ term accounts 
for the $C_4$ symmetry of the surface, and discriminates between the $xy$ and 
$x^2-y^2$ orbitals, as one of them points towards the positively charged Mg 
ions,  reducing the energy of that orbital, whereas the other points towards 
the oxygen atoms. The lowest energy orbital should be the $x^2-y^2$ (if we 
take the oxygen atoms in the $(10)$ and $(01)$ directions).

\section{Density Functional Calculations for Hydrogenated T${\rm i }$ 
\label{sec:DFT}}
In this section we focus on the electronic properties of individual 
Hydrogenated Ti ad-atoms at MgO on-top-of-oxygen, as described with density 
functional theory (DFT) calculations. With this aim, we have employed  
Quantum Espresso\cite{giannozzi2009}, using projected augmented wave 
pseudopotentials, PBE exchange correlation functional and 50-70 Ry of plane 
wave energy cut-off as described elsewhere \cite{giannozzi2009,perdew1998,
blochl1994}. We performed calculations in a structure formed by a bilayer 
of MgO, as shown in Fig. \ref{fig1}, consisting in 36 O atoms (red balls) 
and 36 Mg Atoms (blue balls) together with the hydrogenated Ti (orange ball) 
with one H (green ball). In order to check some results we also performed
a few number of calculations  using a bigger supercell with  64 O atoms, 
64 Mg Atoms and the hydrogenated Ti. The main distortions created by the 
ad-atom in the MgO bilayer are: (i) an upward displacement of the closest 
oxygen(s) to ad-atoms and (ii) a distortion downwards of the Mg atoms located 
below the Ti-bonded oxygen atoms. Our DFT calculations found that a 
hydrogenated Ti atom shows $S=1/2$ \cite{yang2017,willke2018b}. 
The spin density of hydrogenated TiO (with the Ti and H atoms located 
collinear along the $z$ axis) is consistent with a filling of the 
$d_{x^2-y^2}$ orbital, since we are assuming that the Mg atoms first neighbour 
of oxygen are in the $x$ and $y$  axis.

\begin{figure*}[hbt]
\includegraphics[width=0.8\linewidth]{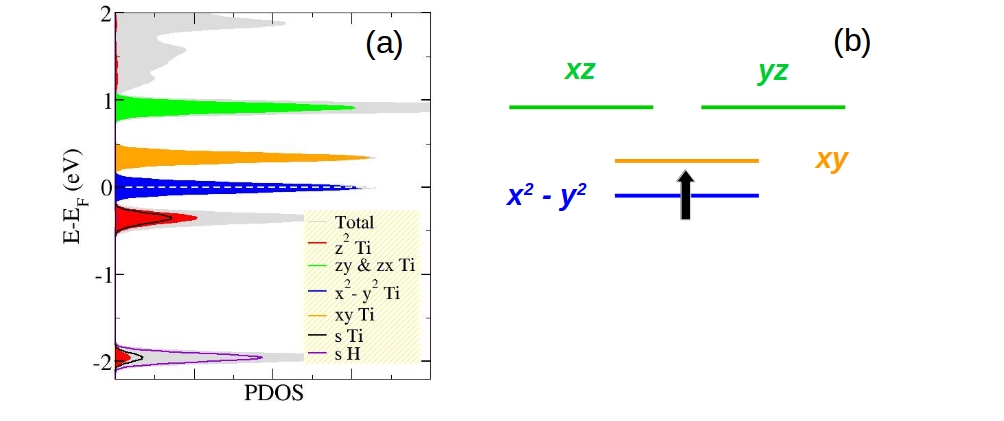}
\caption{ \label{figpdos}(a)  DFT calculations for the  Spin-unpolarized 
Density of States projected over $d-$ and $s-$ orbitals of Ti and the 
$s-$orbital of Hydrogen. The grey shadow shows the total density of 
states. (b) Energy level cartoon.}
\end{figure*}

Fig \ref{figpdos}(a) shows  the projected density of states over $d-$orbitals 
for the hydrogenated Ti at MgO on-top-of-oxygen, computed with no spin 
polarization.  DFT yields  $x^2-y^2$ and a hybrid orbital $z^2-s$ as 
the lowest energy orbitals, within the $d$ manifold.  The  $z^2$ of the 
Ti $d$ shell is strongly hybridized with the hydrogen $1s$ orbital. As a 
result,  the $z^2$ and the $s$ are split in energy and, altogether host 2 
electrons. The $xy$ orbital comes next in energy, and is empty. The orbital 
doublet $xz$ and $yz$ lies higher up in energy. Calculations show that 
$x^2-y^2$ hosts exactly one electron.  It is apparent that  the Hamiltonian 
model Eq. (\ref{H}) with $\ell_z=\pm1,\pm2$ describes the $4$ orbitals $x^2-y^2$, $xy$, $xz$ and $yz$. 

\subsection{Connection between DFT and model Hamiltonian}
\label{DFTsec}
We now explain how to obtain a rough estimate of 
$D$ and $F$ parameters that enter in the crystal field Hamiltonian 
(\ref{H}). The method amounts to fit the energy difference of the peaks in the 
 density of states obtained from a spin-unpolarized DFT calculation
  to those obtained from Eq.  (\ref{H}):
\begin{eqnarray}
E_{x^2-y^2}-E_{xy}&=&48 F \\
2E_{xz}-E_{x^2-y^2}-E_{xy}&=&6|D| 
\label{param}
\end{eqnarray}
Using these equations, from inspection of the density of states  
we infer the values $D\simeq -255$ meV and $F\simeq 7.50$ meV.   
This crude approximation is enough for the scope of this work.  

We can also obtain an  estimates for the modulation of the crystal field 
parameters, $D$ and $F$, as the length of the Ti-O bond is changed from its 
equilibrium position.  The calculation is carried out moving Ti atom and 
relaxing the four closest neighbor Mg atoms, the O atom below and the H 
atom, keeping all the others fixed.  The results of the parameters $D$ and 
$F$, obtained with this procedure allows us to obtain the following relation 
between $D$, $F$ and the strain:
 \begin{eqnarray}
\frac{dF}{dz}&=& -6 \frac{\rm meV}{\AA}
 \label{dFdZ}
 \\
\frac{dD}{dz}&=& +188 \frac{\rm meV}{\AA}
 \label{dDdZ}
 \end{eqnarray}
These values are used later on to estimate how the piezoelectric displacement 
of the Ti-O bond modulates the crystal field values $F$ and $D$, that in turn 
modulate the $g$ tensor. 
  
\subsection{ Calculation of the $g$ tensor from DFT}
We have calculated the $g$ tensor components from our DFT calculations 
using Gauge Including Projector Augmented Waves (GIPAW). GIPAW is a DFT based 
method to calculate magnetic resonance properties \cite{pickard2002}, where 
spin-orbit coupling is implemented in a perturbative way. Our calculations  
for the structure in equilibrium $\delta z=0$, give us a diagonal $g$-tensor 
with components  $g_x=g_y=1.974$ and  $g_z=1.881$.  
As we discuss now, the model Hamiltonian Eq. (\ref{H})  provides  
physical insight on the origin of the anisotropy, and very good agreement 
with the values obtained from DFT. 
      
\section{Calculation of the $g$ tensor from the model \label{sec:gfacmodel}}
We now use the model  Hamitlonian Eq. (\ref{H}) to compute the $g$ tensor.  We do this at two levels of 
approximation. First, we obtain analytical approximate expressions from the 
model Hamiltonian. Second, we obtain the $g$ tensor from the  exact numerical 
solution of the model. Both, the analytical and numerical approach permit to 
relate the $g$ tensor components with  the crystal field parameters $D$ and 
$F$ and the spin orbit coupling $\lambda$. On account of the $C_4$ symmetry of 
the Hamiltonian, the $g$ tensor is diagonal, and has $g_x=g_y$. Therefore,  
we only need to compute $g_z$ and $g_x$.

\subsection{Calculation of $g_z$}
We first consider the response of the electron in the $x^2-y^2$ state to a 
magnetic field in the $z$ direction.   For that matter we need to consider the 
space of 4 states with $\ell_z=\pm 2$ and $S_z=\pm1/2$.  Within this subspace,
spin orbit coupling only acts through the $S_z\ell_z$ term. Therefore,  $S_z$ 
is conserved, and the Hamiltonian for each $S_z$ is given by:
\begin{eqnarray}
{\cal H}_{S_z}(B_z)=
\left(\begin{array}{cc} 
 -\lambda \ell_z S_z +\Delta_- (B_z)    & 24 F   \\
24 F & \lambda \ell_z S_z + \Delta_+ (B_z) 
 \end{array}
 \right)
 \label{h22}
 \end{eqnarray}
 where
$\Delta_{\pm}(B_z)= \mu_B B_z (gS_z\pm\ell_z)$
 with $\ell_z=2$.  Hamiltonian Eq. (\ref{h22}) can be written as:
\begin{eqnarray}
{\cal H}(S_z)=
 g\mu_B B_z S_z + \vec{h}(S_z)\cdot\vec{\sigma}
 \end{eqnarray}
 where 
  \begin{equation}
 \vec{h}(S_z)= \left(24 F,0, -\ell_z(\mu_B B_z+ S_z \lambda) \right)
 \label{vech}
 \end{equation}
We thus have $\epsilon_2(S_z,\pm)=-4|D|+E^{(\pm)}(S_z)$, where:
 \begin{equation}
 E^{(\pm)}(S_z)=g\mu_B B_z S_z \pm 
 \sqrt{(24 F)^2+ \ell_z^2(\mu_B B_z + S_z \lambda)^2}\nonumber
 \end{equation}
We now  Taylor expand the ground state of the $E^{-}$ states 
around $B_z=0$:
\begin{equation}
E^{(-)}(S_z)=-\sqrt{(24 F)^2+\lambda^2}+ g\mu_B S_z  B_z+ \delta g_z   S_z \mu_B B_z
\nonumber
\end{equation}
where
\begin{equation}
g_z=g+\delta g_z=
2 - \frac{4\lambda}{\sqrt{(24 F)^2+ \lambda^2}}
\label{gz}
\end{equation}
Interestingly, there are no  higher order corrections to $\delta g_z$, coming from mixing with the $\ell_z=\pm 1$ manifold. This is confirmed by the comparison of eq. (\ref{gz}) with the results obtained from  exact diagonalization of the complete model   (\ref{H}).  As a result, we can  use equation (\ref{gz}) for obtain the ratio $\frac{F}{\lambda}=1.4$ that gives $g_z$ in agreement with the DFT result $g_z=1.881$    The dependence of $g_z$ on $\lambda$, $F$ and $D$ is shown in Fig. (\ref{fig3}). Note that, as shown in Fig. (\ref{fig3bc}),  $g_z$ does not depend on $D$ and we can  use  $F/\lambda=1.4$ to ensure that $g_z=1.881$.  

\subsection{Calculation of $g_x$}
We now obtain an analytical expression for  $g_x$ for the $x^2-y^2$ ground 
state manifold of Hamiltonian Eq. (\ref{h22}). 
For that matter, we represent the operator $\mu_B B_x (gS_x+\ell_x)$ in the 
basis of eigenstates  of ${\cal H}(\sigma)$:
    \begin{eqnarray}
  |\psi_{-}(\uparrow)\rangle&=& \cos\frac{\alpha}{2}|-2,\uparrow\rangle
  - \sin\frac{\alpha}{2}|+2,\uparrow\rangle \nonumber\\
  |\psi_{-}(\downarrow)\rangle&=& \cos\frac{\alpha}{2}|+2,\downarrow\rangle
  - \sin\frac{\alpha}{2}|-2,\downarrow\rangle 
  \nonumber  
  \label{wave}
  \end{eqnarray}
where the angle $\alpha$ is defined as
  \begin{equation}
 \vec{h}(S_z)= |\vec{h}(S_z)| \left(\sin \alpha(S_z) ,0, \cos\alpha(S_z)
  \right)
 \end{equation}
and $\vec{h}$  is defined in Eq. (\ref{vech}).
In this subspace the matrix elements of $\ell_x$ are zero and the only non 
zero matrix element of $S_x$ reads:
\begin{eqnarray}
\langle \psi^{-}(\uparrow) |S_x  |\psi^{(-)}(\downarrow)\rangle=
-\frac{1}{2}\sin\alpha
\end{eqnarray}

For $B_z=0$, we have
   \begin{eqnarray}
  \sin \alpha= \frac{24 F}{\sqrt{(24 F)^2+ \lambda^2}}  
  \;  \; \;
\cos\alpha=\frac{-|2S_z|\lambda}{\sqrt{(24 F)^2+ \lambda^2}}
 \end{eqnarray}
 
Thus, the eigenvalues of the Hamiltonian are: 
\begin{equation}
\epsilon_{2}(B_x)= -3|D| -\sqrt{(24F)^2+\lambda^2}\pm \frac{g}{2}\mu_B B_x  \sin \alpha
\label{GS}
\end{equation}

We thus have:
\begin{equation}
g_x^{(1)}=  g  \sin\alpha= g\frac{24F}{\sqrt{(24 F)^2+\lambda^2}}
\label{gx0}
\end{equation}

We now consider the contributions to $g_x$ that arise from the virtual transitions to the $\ell=\pm 1$ levels.  These are driven by the combined action of the $\mu_B \ell_x B_x$ and the  flip-flop part of the spin orbit interaction.  This additional contribution gives: 
\begin{equation}
\delta g_x^{(2)} =-\frac{2\lambda}{ 3 |D|+ 24 |F| +|\lambda|}
\end{equation}
so that the $g_x$ factor is given by:
\begin{equation}
g_x=  g\frac{24F}{\sqrt{(24 F)^2+\lambda^2}}-\frac{2\lambda}{ 3 |D|+ 24 |F| +|\lambda|}
\label{gx}
\end{equation}

The anisotropy of the $g$ tensor arises ultimately from the fact that the 
$\ell_z=\pm 2$ states have a strong additional orbital response only when $B$ 
is applied  in the $z$ direction. This extra contribution is quenched by the 
$F$ crystal field term, that leads to states with equal weight on the two 
$\ell_z=\pm2$ states, but promoted by the spin orbit coupling.  The resulting 
anisotropy is thus controlled by the competition between $\lambda$ and $F$.  In addition, $g_x$ has also a contribution that arises from  virtual coupling to the $\ell_z=\pm 1$ states.  For the values of $D,F,\lambda$ adequate to describe Ti-H on MgO,  the dominant contribution to the departe of $g_x$ from the value $g=2$ arises from the virtual coupling to $\ell=\pm 1$. 

So, for $F=0$ we have $g_x^{(1)}=0$, because the spin-orbit coupling correlates 
$S_z$ and $\ell_z$ so that spin flips entail momentum flips, that are 
forbidden, and  $g_z=-2$ because of the dominant orbital contribution. In the 
opposite limit of $\lambda=0$ we recover $g_x=g_z=2$.   If we repeat the 
analysis for $g_y$ we obtain $g_x=g_y$, as expected from the C$_4$ surface's symmetry.

\begin{figure}[hbt]
\includegraphics[width=0.9\linewidth]{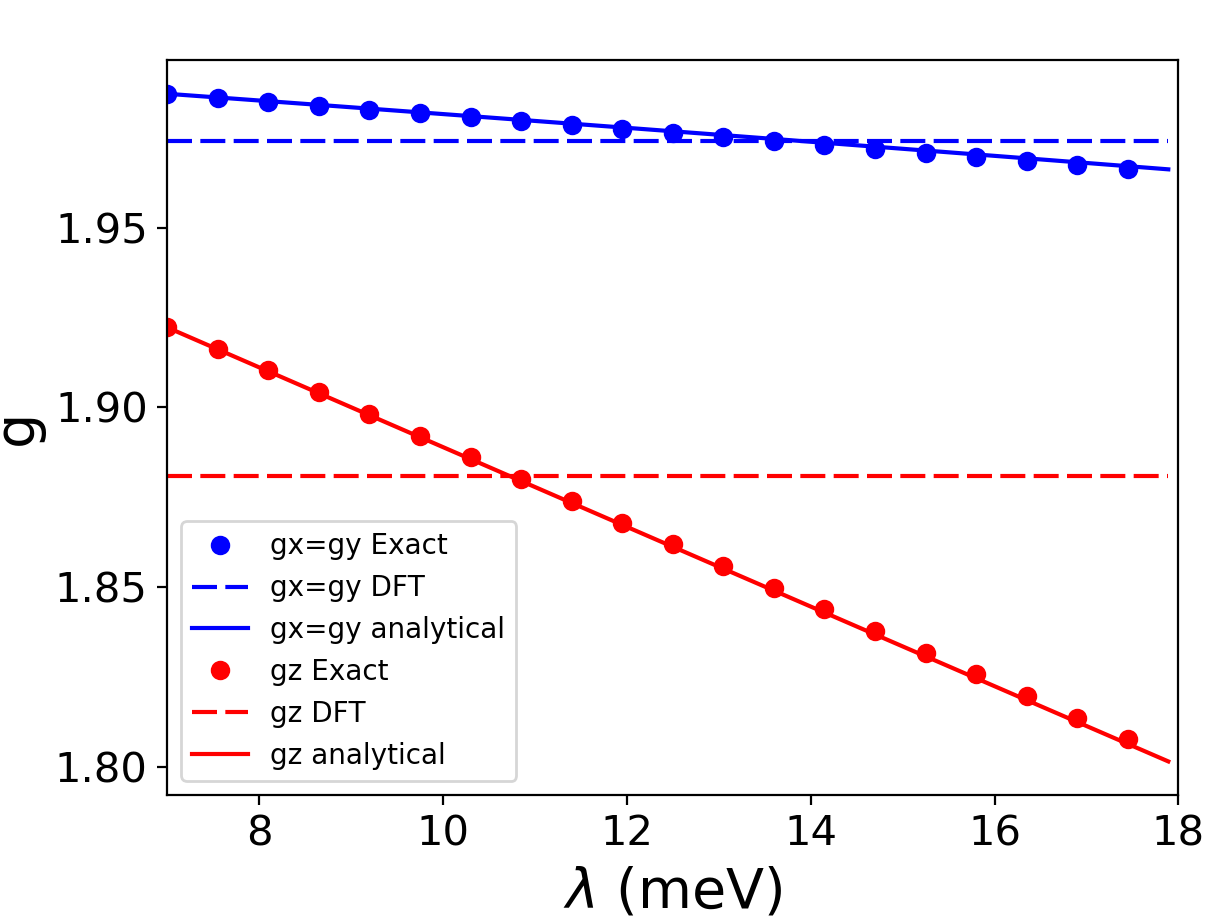}
\caption{ \label{fig3} 
Dependence of $g_x$ and $g_z$ on spin orbit coupling $\lambda$,  for Ti-H on 
oxygen,  obtained in two ways:  solution of full model Eq. (\ref{H}) with 
$D=$-255 meV and $F=14$ meV (symbols) and using analytical results ignoring 
$\ell_z\neq 2$ manifolds (Eq. (\ref{gz}) and Eq. (\ref{gx})) (lines). DFT 
results are shown for reference as dashed lines.}
\end{figure}

\begin{figure}[hbt]
\includegraphics[width=0.9\linewidth]{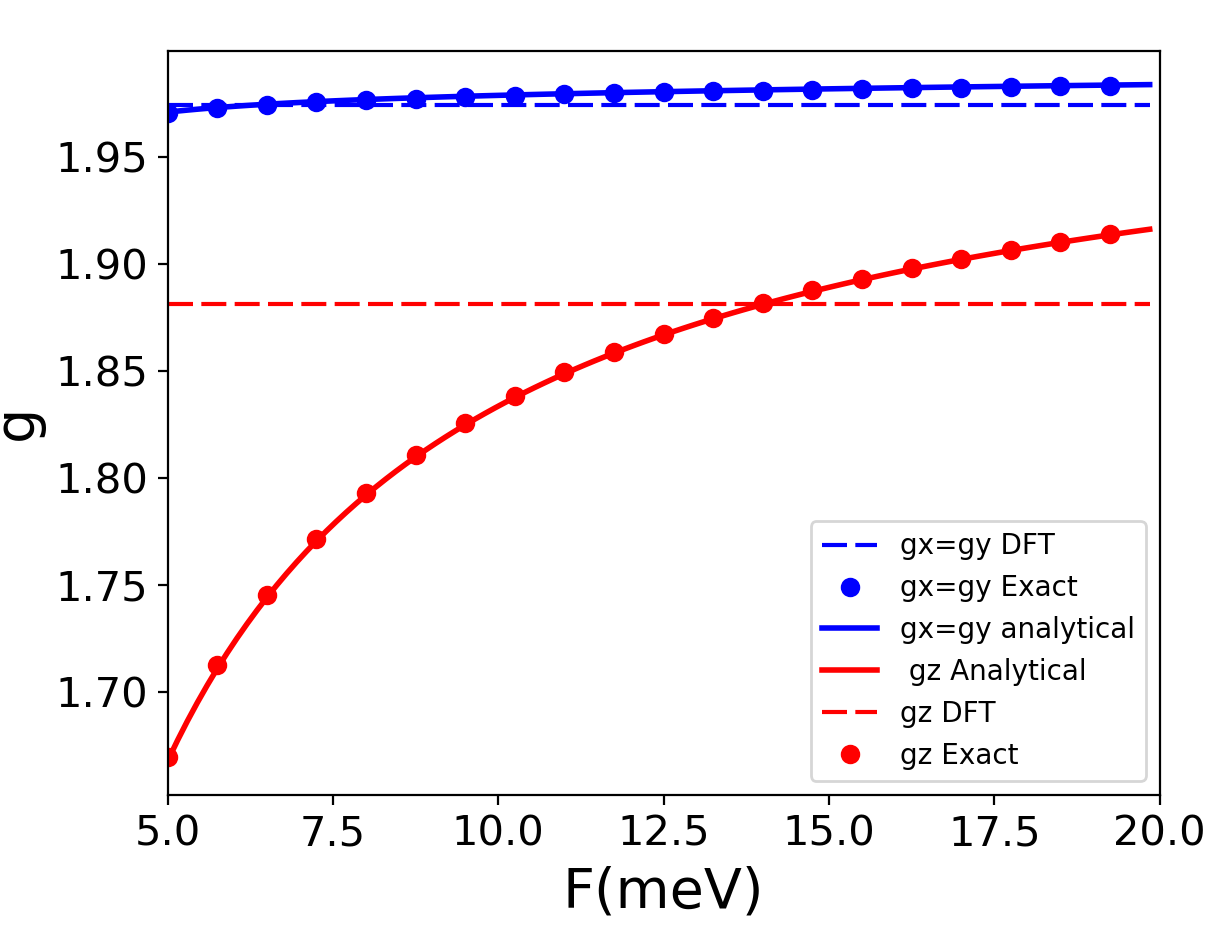}
\includegraphics[width=0.9\linewidth]{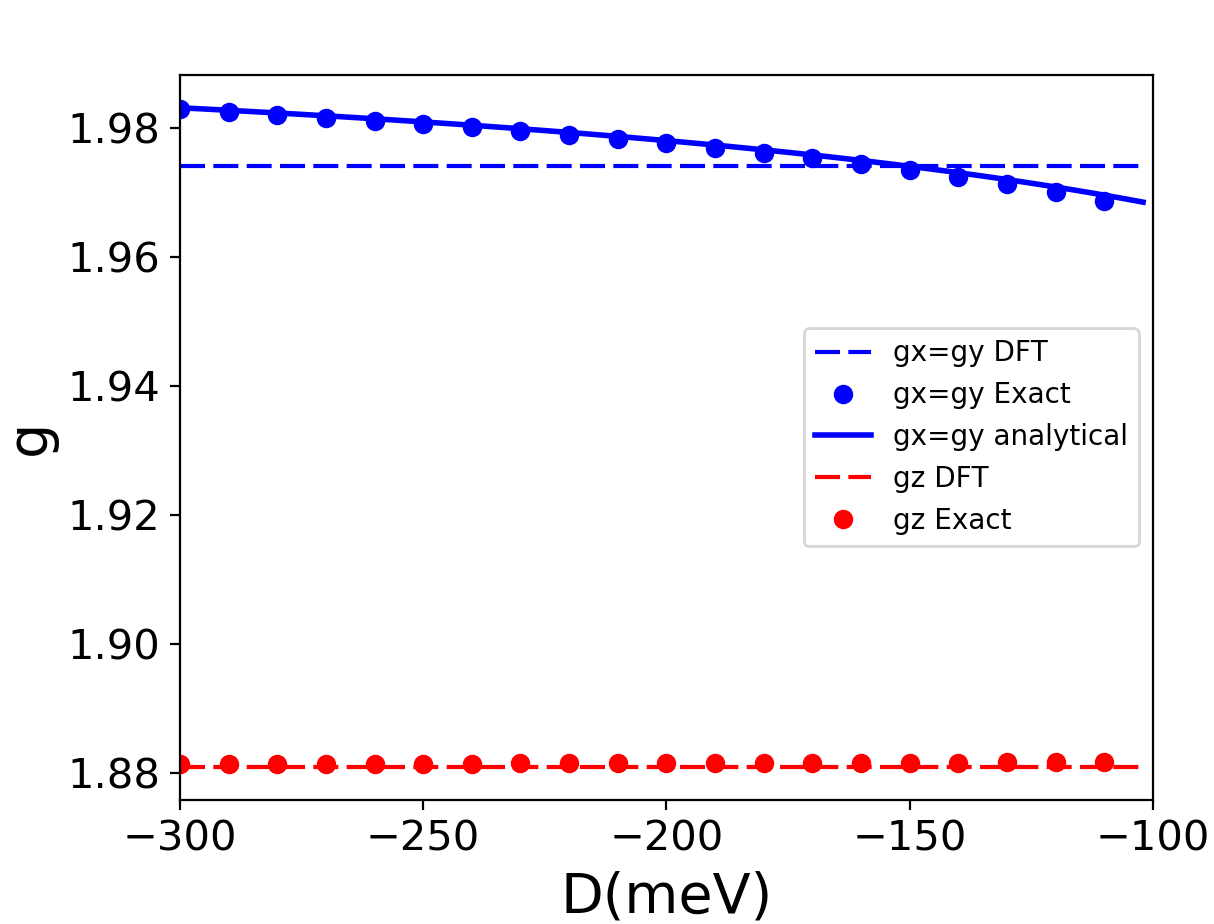}
\caption{ \label{fig3bc} 
Dependence of $g_x$ and $g_z$ on crystal field parameters $F$  (top) and $D$ 
(bottom) for Ti-H on oxygen,  obtained from full model Eq. (\ref{H}) (solid 
points in both panels), DFT (dashed lies) and analytically (solid lines) (Eq. (\ref{gz}) and Eq. (\ref{gx})). 
In both panels we take $\lambda=10$ meV. Top panel: $D=-255$ meV.  
Bottom panel $F=14$ meV.}
\end{figure}

In Fig. \ref{fig3} and Fig. \ref{fig3bc} we show the predictions for $g_x$ 
and $g_z$, as a function of $\lambda$, $D$ and $F$, obtained using both 
the analytical formulas Eq. (\ref{gz}), Eq. (\ref{gx}) and the exact solution 
of  the complete Hamiltonian Eq. (\ref{H}). The DFT results are shown as 
dashed horizontal lines. In Fig. \ref{fig3} and Fig. \ref{fig3bc} (top) we 
take $D=-255$ meV, roughly estimated from DFT, using Eq. (\ref{param}) which 
gives a single particle spectrum in agreement with the results of DFT. In the 
bottom panel of Fig. \ref{fig3bc} we take $\lambda=10$ meV and $F=14$ meV, also inferred 
from comparison with DFT. In Fig. \ref{fig3bc} we fix $\lambda=10$ meV and $F=1.4\lambda$
so that we obtain values very close to those obtained with DFT.
 Finally,  figure \ref{fig3bc} shows that  the 
dependence of $g_x$ on $D$ is small, and $g_z$ does not depend on $D$.
 
Summing up , the results of this section show how, for a model with the 
symmetry adequate for a TiH on top of an oxygen on an $(001)$  MgO surface, 
the $g$ factor is anisotropic, $g_z\neq g_x=g_y$ and how  $g_z$ and $g_x$ 
depend on the crystal field parameters $F$, and to a lesser extent, on $D$. Our analytical model is able to
give $g_x$ and $g_z$ in agreement with the values obtained from DFT.

\section{Piezoelectric modulation of $g$ for T${\rm i }$-H on M${\rm g}$O 
\label{sec:piezog}}
We have shown that a modulation of the $g$ factor anisotropy would induce 
spin-flip transitions (Eq. (\ref {Rabi1})),  and we have computed how the $g$ 
factor components depend on  the crystal field parameters $F$  and $D$.  We 
now argue that an electric field applied perpendicular to the surface of MgO 
modulates $F$ and $D$, and thereby  the $g$ factor anisotropy, resulting in 
spin transitions between the two states of the lowest energy Kramers doublet 
of Eq. (\ref{H}).  
 
Our DFT calculations  show that crystal field parameters $D$ and $F$ are 
functions of the adatom-oxygen distance, $z$:  $D(z)$, $F(z)$ (see Eq. 
(\ref{dFdZ}) and Eq. (\ref{dDdZ})). 
We denote the equilibrium 
position  by $z_{\rm eq}$. 
The  electric 
field across the gap between the STM and the MgO surface, $E= V_{tip}/d$, 
where $d$ is the tip-MgO distance,   induces a force on the adatom, 
$F=q_{\rm adatom} V_{\rm tip}/d$ on account of its charge $q_{\rm adatom}$
t\cite{Baumann2015,lado2017,yang2019}. 
This force is compensated by a restoring elastic force $F= -k \delta z$. Thus, 
the  adatom equilibrium position is displaced 
  by\cite{lado2017}:
 \begin{equation}
 \delta z(t)=  \frac{q_{\rm adatom}V_{\rm tip}(t)}{kd}
 \label{dz}
 \end{equation}
This equation is valid for a time dependent $V_{\rm tip}$ as long as its 
Fourier components are away from the mechanical  resonance frequency of the 
stretching mode, $\sqrt{\frac{k}{M}}$, where $M$ is the mass of the adatom.  
According to our DFT calculations\cite{lado2017,yang2019}, this frequency  is 
up in the THz range, as long as we ignore the contributions coming from the 
off-plane (flexural) phonons of the MgO. In the following we assume  
$V_{\rm tip}(t)=V^0_{RF} \cos(2\pi f t)$ so that we have 
 \begin{equation}
  \delta z(t)= \frac{q_{\rm adatom}V^0_{\rm RF}(t)}{kd}\cos(2\pi f t)
  \equiv \delta z_0 \cos(2\pi f t)
  \label{dz2}
 \end{equation}

\noindent From our DFT calculations for Ti-H on MgO\cite{yang2019} we 
obtain $k=290 $ eV nm$^{-2}$, so that for RF tip voltages values ranging from 
$eV^0_{RF} =10$ meV to $eV^0_{RF} =20$ meV and $d=5 \AA$, the piezoelectric 
displacement amplitude goes from $\delta z_0 =0.07\,$ pm to 
$\delta z_0 =0.14\,$ pm.

The modulation of crystal field parameters  $F$ and $D$  with the Ti-O bond 
length leads to a modulation of the $g$ tensor: 
\begin{equation}
\delta g_a=\left( \frac{\partial g_a}{\partial F} \frac{\partial F}{\partial z}
+ \frac{\partial g_a}{\partial D} \frac{\partial D}{\partial z}
\right)
 \delta z(t)
 \label{deltag}
\end{equation}
It must be noted that this equation is also valid in the  $DC$ limit.

We now proceed to estimate the magnitude of the Rabi coupling associated to 
the modulation of the $g$ factor.  We do that using two different methods 
that, as we discuss below, give the same result. The first method consist on 
using Eq. (\ref{Rabi1}).  In the second method we evaluate directly the matrix 
elements of the crystal field operators, using the same approach used in 
previous works\cite{Baumann2015,lado2017}.

\subsection{Rabi coupling from the $g$ factor anisotropy}
In order to compute the Rabi coupling from eq. (\ref{Rabi1}), we need to 
compute eq. (\ref{deltag}).  
For that matter, we obtain $ \frac{\partial g_a}{\partial F}$ and 
$\frac{\partial g_a}{\partial D}$  from our model Hamiltonian, and we 
use $\frac{\partial F}{\partial z}, \frac{\partial D}{\partial z}$ calculated
from DFT calculations in Sec. (\ref{DFTsec}).

\begin{figure}[hbt]
\includegraphics[width=1.\linewidth]{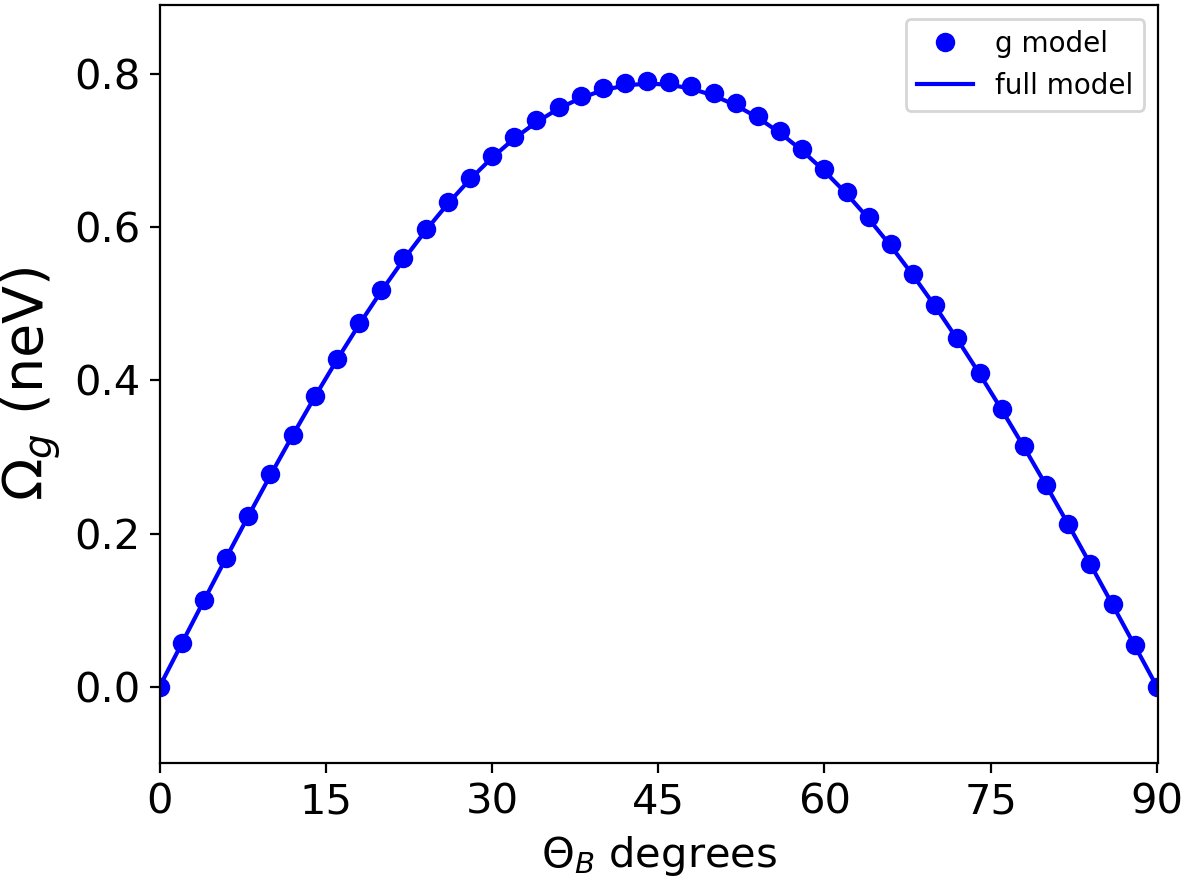}
\includegraphics[width=1.\linewidth]{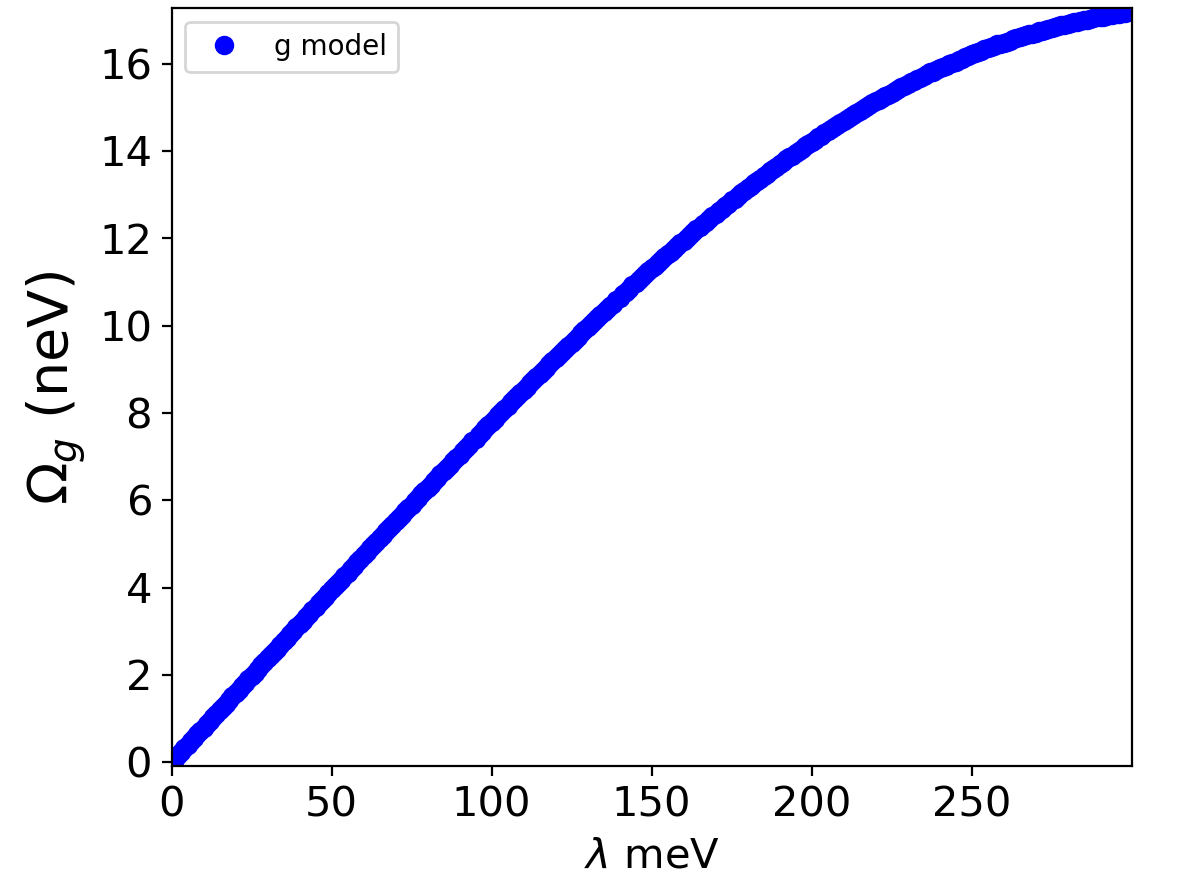}
\caption{ \label{fig4} Top:  Rabi coupling due to $g$ factor modulation as a 
function of the angle $\theta_B$ between $\vec{B}$ and surface calculated with 
two methods: Eq. (\ref{Rabi1}) and Eq. (\ref{rabiCF}). $\lambda=10$ meV, 
$F=1.4 \lambda=14$ meV, $D=-255$ meV, $B=1T$, $V_{ F}=20$meV, STM-surface 
distance  $d=5\AA$, $k=290$eV$/{\rm nm}^2$. Bottom: Dependence of $\Omega_g$ 
on $\lambda$, keeping all the other constants the same and $\theta_B=45$ 
degrees.}
\end{figure}

We are now in position to estimate $\Omega_g$, combing Eq. (\ref{Rabi1}), 
Eq. (\ref{dFdZ}), Eq. (\ref{dz2}) and  Eq. (\ref{deltag}).   
We now take $V^0_{RF}=20$ meV,  $d=5\AA$ and $k=290$ev$/{\rm nm}^2$ 
(taken from DFT calculations). This yields a strain of the $Ti-O$ distance 
of $\delta z_0= 0.14$pm. 
In Fig. (\ref{fig4}) we plot the magnitude of the Rabi coupling so obtained, 
as a function of the  angle between the magnetic field and the surface, 
$\theta_B$ for $B=1$ Tesla. The first thing to note is that the magnitude of 
$\Omega_g$ is between one and two orders of magnitude smaller than the 
experimental  values reported in our previous work\cite{yang2019}. Therefore,
other mechanism, most likely exchange modulation\cite{lado2017},   
has to be involved in  the electric field driving of the spin for 
ESR-STM  for Ti-H/MgO.
 
The magnitude of $\Omega_g$ scales linearly both with the applied field $B$ 
and with the RF electric field $\frac{V_{RF}}{d}$.  The optimal angle to 
maximize $\Omega_g$ is close to 45 degrees.  In contrast, the exchange 
mechanism is independent of $B$ and scales exponentially\cite{yang2019} with 
$d$. Whereas the $g$ factor modulation is not dominant for Ti-H on MgO,  it 
could be the dominant factor in heavier adatoms.  To show this, in the bottom 
panel of Fig. \ref{fig4} we plot $\Omega_g$  ramping $\lambda$, keeping all 
the other parameters the same and taking $\theta_B=45$.
It is apparent that, for a wide range,  $\Omega_g$ scales linearly with spin 
orbit coupling.  Expectedly, $\Omega_g$ vanishes for $\lambda=0$, as the $g$ 
factor anisotropy is driven by $\lambda$.

\subsection{Rabi coupling from crystal field matrix elements}
We now carry out a sanity check.  Given that the electronic spin flip 
transitions described by the effective $S=1/2$ model  of Eq. (\ref{h00}) and Eq.
(\ref{V}), arise ultimately from the modulation of the crystal field operators 
of the parent Hamiltonian of Eq. (\ref{H}), we have computed the Rabi coupling 
using the parent model as well. To do so, we first obtain the two lowest 
eigenstates of Hamiltonian from Eq. (\ref{H}), 
${\cal H}|\pm\rangle=\left(E_g \pm \frac{\Delta_z}{2}\right)|\pm\rangle$ and 
we then compute the matrix elements of the perturbation operator:
\begin{equation}
{\cal V}(t)= \frac{\partial D}{\partial z}\delta z(t)\ell_z^2 
+ \frac{\partial F}{\partial z}\delta z(t) 
	\left((\ell^{(+)})^4+(\ell^{(-)})^4\right)
\end{equation}

\noindent where the time dependence is described by Eq. (\ref{dz2}).
We thus define:
 \begin{eqnarray}
 \Omega_{CF}&=&\frac{\partial D}{\partial z}\delta z_0 \langle + |\ell_z^2 |
-\rangle +\nonumber\\ 
	 &+&\frac{\partial F}{\partial z}\delta z_0  \langle + |\left((\ell^{(+)})^4+(\ell^{(-)})^4\right)
 |-\rangle 
 \label{rabiCF}
 \end{eqnarray}
 
The results of the calculation of $ \Omega_{CF}$ as function of $\theta_B$, 
obtained with the same values of $\lambda$, $D$, $F$, $d$, $V_{RF}$, $k$, and 
$B$  of the previous subsection,  are shown in Fig. \ref{fig4} and, 
expectedly, are in full agreement with those obtained using Eq. 
(\ref{Rabi1}). This agreement validates our analysis.

\section{$g$ factor modulation of   F${\rm e}$ on M${\rm g }$O \label{sec:Fe}}
The results of the last paragraph  show that it is possible to interpret 
the modulation spin-driving coming from the modulation of the crystal field 
parameters (Eq. (\ref{rabiCF})) in terms of a modulation of the $g$ 
factor (Eq. \ref{Rabi1}).    In the seminal work of Baumann {\em et al.}
\cite{baumann2015electron}, the modulation of the crystal field (CF) was 
proposed as the driving mechanism for ESR-STM of Fe/MgO.  Here we address the 
question of whether  we can  recast the CF mechanism in terms of the $g$ 
factor modulation, for the case of Fe on MgO as well. 

In order to find the answer, that turns out to be negative, we need to model 
the $g$ factor modulation for Fe on MgO and to compare with the results 
obtained from the CF modulation. The main difference with the case of Ti-H is 
that the ground state of Fe on MgO has $S=2$. Therefore, a multi-electronic 
description is necessary\cite{Baumann2015,Baumann_Donati_prl_2015,lado2017}. 

We follow our own work \cite{lado2017} and we model Fe on MgO with a two 
levels of complexity. First, a microscopic Hamiltonian for 6 electrons  in 
the $d$ orbitals of Fe,  in the presence of a crystal field, spin orbit 
coupling, Coulomb interaction and Zeeman interaction:
  \begin{equation}
  {\cal H}_{\rm Fe}=H_{\rm CF}+H_{\rm SOC}+ H_{\rm Z} + V_{ee}
  \label{CI}
  \end{equation}
The single particle crystal field Hamiltonian  reads:
\begin{equation}
{\cal H}_{CF}= D\ell_z^2 +F\left(\ell_{x}^4+\ell_{y}^4\right) 
\label{CFFe}
\end{equation}
As explained in the appendix (\ref{CFAPP}),  this CF Hamiltonian turns is 
almost identical to the one we have used in Eq. (\ref{H})  Ti-H/MgO.   As we 
did in the case of Ti-H/MgO, we can infer $D$ and $F$ from DFT calculations
\cite{lado2017}. For $z=z_{eq}$ we obtain, from DFT calculations, $F=-10$ meV 
and $D=-290$ meV \cite{lado2017}. The spin orbit coupling constant for Fe is 
$\lambda=35$ meV. \cite{lado2017}

The many-body Hamiltonian can be solved exactly, by numerical diagonalization 
in a space made with all the states that accommodate 6 electrons in 5 spin 
degenerate $d$ orbitals. The lowest energy manifold has 5 states, 
corresponding to a ground state with $S=2$ and can be described in terms of an 
effective spin model: 
 \begin{eqnarray}
H_{\rm eff}&=& -{\cal D}_2 S^2_z + {\cal D}_4 S^4_z  - 
{\cal F}(S^4_+ + S^{4}_-) \nonumber \\  
&+&\mu_B {\bf B} \cdot g \cdot  {\bf S} 
\label{hfe1}
\end{eqnarray}
where the spin  operators act on the $S=2$ subspace. The main difference with 
the $S=1/2$ case is the presence of single ion anisotropy terms. The anisotropy
terms ${\cal D}_2,{\cal D}_4, {\cal F}$  and the $g$ tensor can be obtained 
from the diagonalization of Hamiltonian Eq. (\ref{CI}). We obtain
${\cal D}_2=4.9$ meV, ${\cal D}_4=0.23$ meV and ${\cal F}=11$ $n$eV. With 
these numbers, the spectrum of the $S=2$ manifold has a ESR active 
space formed by a doublet of states with $S_z=\pm 2$, that we denote as 
$|0\rangle$ and $|1\rangle$. Yet, this doublet is fundamentally different
\cite{hoffman1994} from 
the $S=1/2$ Kramers pair, as it has a zero field splitting, given by $\Delta=48 {\cal F}=0.5 \mu$eV,
 due to quantum 
spin tunneling\cite{Klein_ajp_1951,garg1993topologically,
Wernsdorfer_Sessoli_science_1999,Delgado_Loth_epl_2015}.  Thus, the ESR active 
doublet for Fe on MgO  can not be described in terms of a Zeeman only 
Hamiltonian.  At $B=0$, none of the two lowest energy states  has a magnetic moment\cite{Delgado_Loth_epl_2015}.  However, application of a modest off-plane field is enough to induce an off-plane magnetic moment in the two lowest energy states, on account of the small value of $\Delta$.

Diagonalizations of the multi-electronic Hamiltonian Eq. (\ref{CI}) at finite 
magnetic field permit to derive the $g$ tensor. Expectedly for a system with 
$C_4$ symmetry, it is diagonal in the cartesian basis.  The values of the $g$ 
tensor do depend on the single particle crystal field parameters $D$ and most 
notably on $F$ and $\lambda$.  For the values quoted above, we obtain 
$g_z=2.8$ and $g_x=g_y=2.0$. 

Importantly, all the constants in the effective Hamiltonian Eq. (\ref{hfe1}) 
do depend strongly on the single particle crystal field parameter $F$, that in 
turns depends on the piezoelectric displacement\cite{lado2017}. Following a 
similar argument that the one used for $d^1$  atoms, we can calculate the Rabi 
frequency derived from the effective Hamiltonian.  We break it down in two 
types of terms:
\begin{equation}
\Omega_{eff} = \Omega_{ZFS} + \Omega_g
\end{equation}
The first comes from the modulation of the zero field  energy constants, 
${\cal D}_2$, ${\cal D}_4$, and ${\cal F}$ and was absent in the case of 
$S=1/2$ adatoms . The dominant contributions\cite{lado2017} arise from the 
modulation of the $F$ term  in the single particle crystal field of Eq. 
(\ref{CFFe}):
\begin{eqnarray}
&&\Omega_{ZFS} = \left(- \frac{\partial {\cal D}_2 }{\partial F} \langle 0  |S^2_z|1 
\rangle\right.\nonumber \\
&&\left. +\frac{\partial {\cal D}_4 }{\partial F} \langle 0  |S^4_z|1 \rangle  - \frac{\partial {\cal F} }{\partial F} \langle 0 | S^4_+ + S^4_-|1 \rangle \right) \frac{\partial F }{\partial z}
\label{ZFS}
\end{eqnarray}
where $\frac{\partial F }{\partial z}=280 $meV $/nm$,  obtained from DFT in 
a previous publication\cite{lado2017}.
The second  class of contribution to the Rabi coupling comes from the $g$ 
factor modulation, very much like the $S=1/2$ case:
\begin{eqnarray}
\Omega_{g} =\mu_B \left(\frac{\partial g_z }{\partial F}   B_z \langle 0 |S_z| 1 \rangle + \frac{\partial g_x }{\partial F}  B_x \langle 0 |S_x| 1 \rangle \right)  \frac{\partial F }{\partial z}
\label{h1}
\end{eqnarray}

\begin{figure}[hbt]
\includegraphics[width=0.85\linewidth]{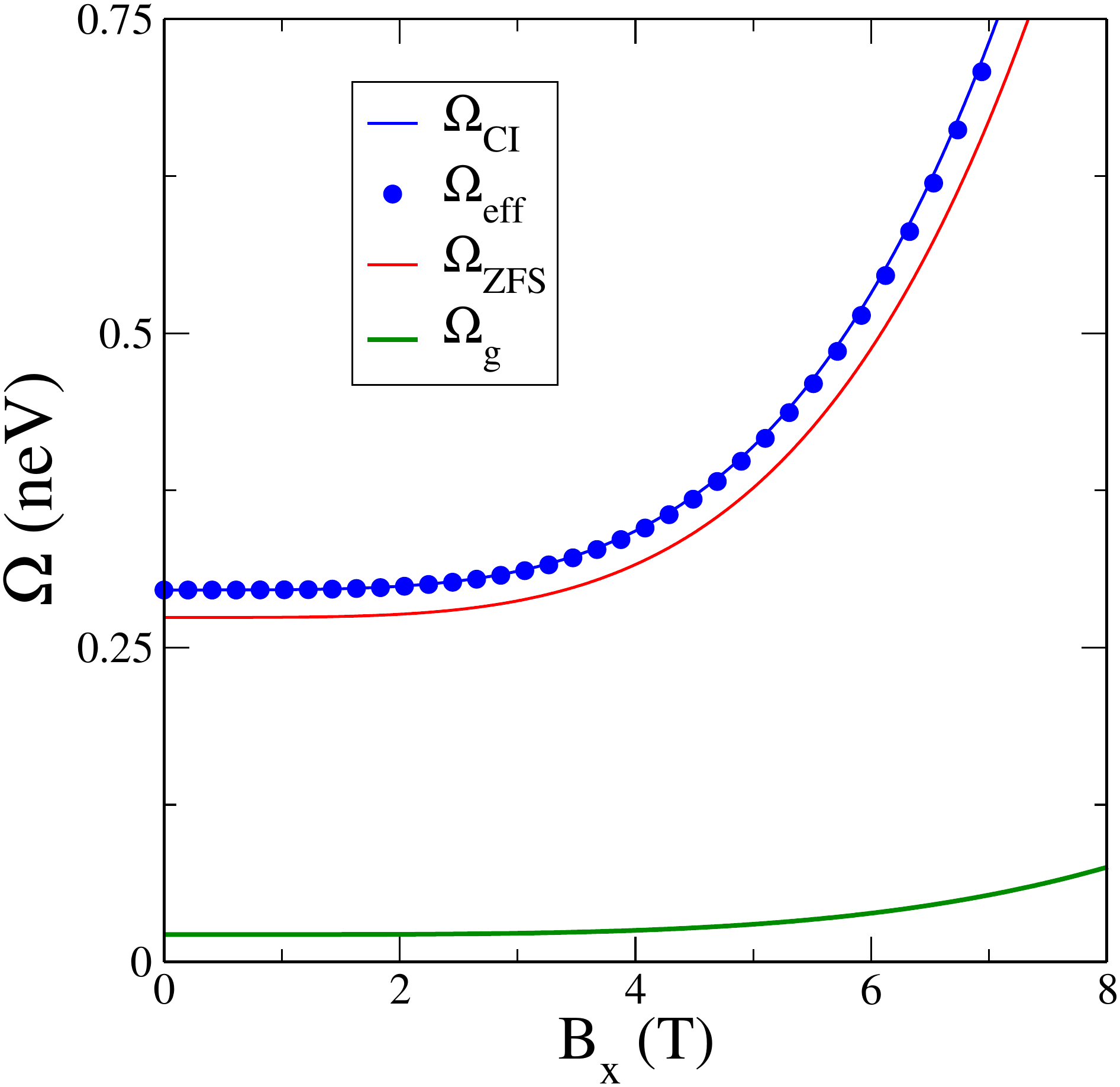}
\caption{ \label{figfe2}.  Breakdown of Rabi coupling for Fe on MgO as a 
function of  in plane magnetic field $B_x$. The calculation were performed
for $B_z=0.2$ T , $k=600$ eV$/$nm$^2$, $d=0.6$ nm, $V_{RF}=8$ mV, $D=-290$ meV,
$F=-10$ meV, $\lambda=35$ meV and $q=2e$}
\end{figure}

We can assess the relative contribution of the zero field splitting  and the 
$g$ tensor modulations in the following way. We first compute the Rabi 
coupling using the whole multi-electron Hamiltonian\cite{lado2017} and we refer 
to this as $\Omega_{CI}$.  The calculation, done for $B_z=0.2$ T , 
$k=600$ eV$/$nm$^2$, $d=0.6$ nm, $V_{RF}=8$ mV, $D=-290$ meV, $F=-10$ meV, 
$\lambda=35$ meV and $q=2e$ is shown in Fig. (\ref{figfe2}) as a function of 
the in-plane field $B_x$, together with the different contributions, 
$\Omega_{ZFS}$ and $\Omega_{g}$, computed using the effective Hamiltonian Eq. 
(\ref{hfe1}), Eq. (\ref{ZFS}) and Eq. (\ref{h1}). It is apparent that the 
$\Omega_{CI}=\Omega_{ZFS}+\Omega_{g}$, which validates our methods.  

Importantly, our calculations show that the modulation of the $g$ factor is 
not a dominant contribution to the spin transitions driven by the modulation 
of the crystal field parameter $F$ due to off-plane piezoelectric distortion 
of the adatom.  In addition, it was found in Ref. \onlinecite{lado2017} that 
the exchange modulation mechanism is probably dominant for Fe. Therefore, the 
$g$ factor modulation plays a marginal role in the case of Fe.

\section{Modulation of the  hyperfine interaction\label{sec:hyper}}
Here we briefly address how the modulation of the $g$ factor anisotropy 
affects the hyperfine interaction.  Recently,   electrical control of an 
individual nuclear spin  of $S=1/2$ Cu atom was demonstrated using  STM-EPR 
\cite{yang2018}. For simplicity here we consider the case of Ti-H on MgO, for 
which hyperfine splittings have been observed  experimentally 
\cite{willke2018b}. We consider a simplified hyperfine model where only the 
contact interaction term is considered:
\begin{equation}
{\cal V}_{\rm HF}=A{\bf I}\cdot{\bf S} 
\label{h1hf}
\end{equation}
For simplicity, the dipolar and quadruple terms are neglected, although they 
are known to be relevant for Ti-H on MgO\cite{willke2018b}.  We now address 
how the modulation of the $g$ factor anisotropy affects this Hamiltonian and  
we find that the effective hyperfine coupling becomes anisotropic. 

For that matter, we consider the representation of the isotropic hyperfine 
operator Eq. (\ref{h1hf}) in the basis set defined by the tensor product of 
the lowest energy  eigenstates of Hamiltonian Eq. (\ref{H}), whose wave 
functions are given  by  Eq. (\ref{wave}), and the eigenstates of the 
nuclear spin operator $I_z$.  The resulting Hamiltonian  reads: 
\begin{equation}
{\cal H}_{HF}= A_{||} I_z S_z + A_{\perp} \left(I_x S_x + A_y I_y S_y\right)
\label{h2hf}
\end{equation}
where 
\begin{equation}
A_{||} = A, \;\; A_{\perp}=A\sin\alpha =  A \frac{24F}{\sqrt{(24F)^2 + 
\lambda^2}} 
\end{equation}
where $F$ and $\lambda$ are the crystal field and spin orbit coupling in Eq. 
(\ref{H}). It is apparent that the modification of the hyperfine interaction 
is connected to the $g$ factor anisotropy. 

We now discuss how the modulation of the $g$ factor could lead to nuclear 
spin-transitions that preserve the electronic spin. We consider the situation 
where a magnetic field induces an electronic Zeeman  splitting so that  the 
eigen-states of the electro-nuclear Hamiltonian can be split in  two groups, 
according to their electronic spin\cite{yang2018}. We are interested in 
transitions between the low energy manifold, so that initial and final state 
belong to the low energy group, and only the nuclear spin changes in the 
transition. We thus consider transitions between two eigen-states that differ 
by single nuclear spin flip\cite{yang2018}:
\begin{equation}
|I_z\rangle_-= \sqrt{1-\epsilon^2} |\downarrow\rangle  |I_z\rangle +
\epsilon  |\uparrow\rangle |I_z-1\rangle 
\label{nuc1}
\end{equation}
and 
\begin{equation}
| I_z-1 \rangle_-=   \sqrt{1-\epsilon^2}|\downarrow\rangle  |I_z-1\rangle +
\epsilon  |\uparrow\rangle  |I_z-2\rangle 
\label{nuc2}
\end{equation}
where $\epsilon\propto \frac{A_{\perp}}{g_z\mu_B B_z}<<1$. These states have 
a  dominant $\downarrow$ electronic spin component and  a small mixing due to 
the non-resonant  spin-flip hyperfine interaction. 

It is apparent that a perturbation that flips the electronic spin can induce 
transitions between these two states: 
\begin{equation}
_-\langle I_z-1 | S^{-}|I_z\rangle_- \propto \epsilon\sqrt{1-\epsilon^2} \simeq \frac{A_{\perp}}{g_z\mu_B B_z}
\end{equation}
showing the electronic driven nuclear spin transition matrix element is 
proportional to  the hyperfine interaction. 
Thus, the same modulation of the g-tensor that drives electronic spin 
transitions, when $f$ in the range of the electronic Zeeman transition, can 
also drive nuclear-spin flip transitions if $f$ is in the range of the 
hyperfine interaction, as shown experimentally in the case of Cu on 
MgO\cite{yang2018}. 

We finally note that  the modulation of $F$ will in turn change $A_{\perp}$ 
providing a time dependent electron-nuclear perturbation,
\begin{equation}
\delta{\cal V}_{HF}= \delta A_{\perp}(t)  \left(I_x S_x + I_y S_y\right)
\label{h2hf}
\end{equation}
where $\delta A_{\perp}= \frac{\partial A_{\perp}}{\partial F}
\frac{\partial F}{\partial z} \delta z(t)$. However, this  electron-nuclear 
flip-flop operator can not mix the states  (\ref{nuc1}) and (\ref{nuc2}).  The 
flip-flip modulation can induce EPR like transitions, between state 
(\ref{nuc1}) and
\begin{equation}
|I_z-1\rangle_+= \sqrt{1-\epsilon^2} |\uparrow\rangle  |I_z-1\rangle +
\epsilon  |\downarrow\rangle |I_z\rangle 
\label{nuc3}
\end{equation}
This could be a relevant mechanism for ESR-STM in systems with very 
large hyperfine interaction, such as Bi in silicon
\cite{George_Witzel_prl_2010} or perhaps Cu/MgO \cite{yang2019}.  
Hyperfine driven electric spin dipole resonances have been reported in 
semiconductor quantum dots\cite{Shafiei2013}.

\section{Role of anisotropic $g$ factor and the exchange driven mechanism for EPR-STM
 \label{sec:exch}}
  
Although the main scope of this paper is to propose a new mechanism for the 
electric field driving of the surface spins in STM-ESR,   we briefly 
comment here on the role that the $g$ factor anisotropy plays on the 
exchange-modulation mechanism that we proposed in Ref. \onlinecite{lado2017} 
and has been experimentally observed\cite{yang2019} for Ti-H adatoms on MgO.  
We now consider a Hamiltonian for the surface spin that, in addition to the  
Zeeman term, given by Eq. (\ref{h00}), has also  exchange interaction with the 
tip. The magnetic moment of the tip is described semiclassically\cite{lado2017,
yang2019}, so that the Hamiltonian for the surface spin reads:
  \begin{equation}
  {\cal H}_{\rm ex}={\cal H}_Z + J(z) \vec{n}_T\cdot\vec{S}
  \end{equation}
where (see Fig. \ref{fig1})
\begin{equation}
\vec{n}_T= \left(\cos(\theta_B+\delta),0,\sin(\theta_B+\delta)\right)
\end{equation}
describes the orientation of the tip moment,
$\vec{B}=B_0 \left(\cos(\theta_B),0,\sin(\theta_B)\right)$ is the external
 magnetic field forming and angle $\theta_B$ with the MgO surface as shown
in Fig. \ref{fig1} and $J(z)$ is the tip-adatom exchange interaction, 
that depends on the tip-surface distance $z$. 

In the appendix \ref{rabiJsec} we derive an expression for the Rabi energy
associated to the modulation $\delta J$ of the exchange  formula:
\begin{eqnarray}
\Omega_J={\cal E}_J \left(\frac{\Delta g_x}{g} -\frac{\Delta g_z}{g}\right)
 \cos\theta_B\sin\theta_B \cos\delta+\nonumber\\
+{\cal E}_J\left(1+\frac{\Delta g_x}{g} \cos^2\theta_B+\frac{\Delta g_z}{g}\sin^2\theta_B
\right)
\sin\delta
\end{eqnarray}
where  we write the ansotropic $g$ factor as 
\begin{eqnarray}
g_x= g +\Delta g_x
\nonumber \\
g_z= g +\Delta g_z
\end{eqnarray}
where $\Delta g_x$ and $\Delta g_z$ are the  static  contributions to the 
$g$ factor anisotropy and
\begin{equation}
{\cal E}_J\equiv \frac{\delta J}{2\Delta_z^*} g\mu_B B
\end{equation}
and
\begin{equation}
\Delta_Z^*\equiv \sqrt{ (g_x \mu_B B_x+ Jn_x)^2 + (g_z\mu_B B_z+J n_z)^2}
\end{equation}
Let us consider now two different limits for this complicated formula. 
We study first  the case  $\delta=0$, {\em i.e.}, when the tip magnetic moment 
is aligned with the external magnetic field. This amounts to assume that the 
tip magnetic moment has an isotropic $g$ factor. For $\delta=0$   the exchange 
modulation Rabi splitting reads:
\begin{eqnarray}
\Omega_J=\frac{{\cal E}_J}{2} 
 \sin2\theta_B 
\left(\frac{\Delta g_x}{g} -\frac{\Delta g_z}{g}\right)
\label{rabiJ-aniso}
 \end{eqnarray}
Thus,  this equation makes it  apparent that the $g$ factor anisotropy of the 
surface spin is essential if the tip spin is aligned with $B$ ($\delta=0$). We 
note that, in spite of the similar aspect of Eq. (\ref{Rabi1}) and Eq. 
(\ref{rabiJ-aniso}), they describe different mechanisms. In Eq. (\ref{Rabi1}), 
the transitions are driven by the anisotropic modulation $\delta g$ of the $g$ 
factor anisotropy. In Eq. (\ref{rabiJ-aniso}),  the transitions are driven by 
the modulation of the tip-surface spin $\delta J$, but are enabled by the 
static  $g$ factor anisotropy, given by $\Delta g$. 
 
We now consider the case  $\Delta g_x=\Delta g_z=\Delta g$,  {\em i.e.}, when 
the surface spin has an isotropic $g$ factor. We  obtain: 
\begin{eqnarray}
\Omega_J={\cal E}_J \left(1+ \frac{\Delta g}{g}\right)\sin \delta
\label{anisotip}
\end{eqnarray}
Thus, the exchange-modulation Rabi depends on the misalignment angle 
between the tip moment and the applied field. 
In both cases, the exchange modulation Rabi splitting requires that either the 
tip or the surface spin, or both, have to be misaligned with respect to the 
applied field. 
  
\section{Electric control of  the resonance frequency \label{sec:shift}}

In the ESR-STM experiments there is a DC bias, with amplitude $V_{DC}$,  
super-imposed to the AC bias. In this section we consider the shift of the 
resonance frequency of a $S=1/2$ adatom with anisotropic $g$ factor on account 
of the DC  electric field between the tip and the surface. We consider the 
case of Ti-H/MgO.  The underlying mechanism is the same that gives rise to the 
spin transitions: application of an off-plane electric field  induces a 
strain $\delta z$ of the bond between the Ti adatom and the  oxygen 
atom  underneath(see Eq. (\ref{dz})). This leads to a modulation of the crystal field 
parameters $D$ and $F$ that in turn shifts the $g$ tensor.   

In the case of the g-factor modulation, we can obtain an expression for the 
shift of the resonance frequency for a given DC modulation $\delta g_x$ and 
$\delta g_z$ of the the $g$ tensor, up to linear order in $\Delta g_{a}$:
\begin{equation}
\hbar \delta \omega\simeq\mu_B^2\frac{ g_x\delta g_x B_x^2
+g_z \delta g_z B_z^2}{\Delta_z} 
\label{shift:eq}
\end{equation}
where $\Delta_z$ is the unperturbed modulation. We emphasize that $\delta g_x$ 
and $\delta g_z$ in Eq. (\ref{shift:eq}) are the time independent contributions
to the $g$ factor anisotropy that arise from application of a DC electric field. 
\begin{figure}[t]
\includegraphics[width=1.\linewidth]{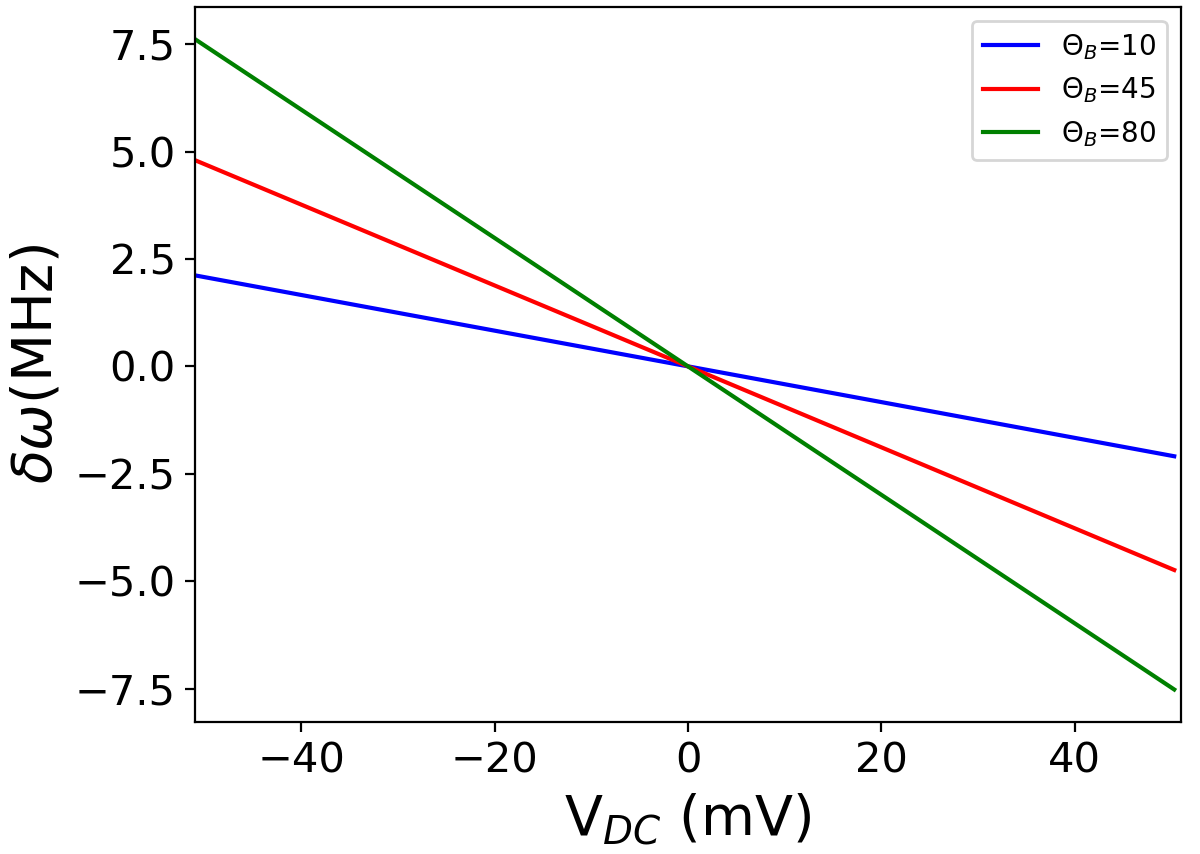}
\caption{ \label{fig7}.  Shift of the resonance frequency for
$B=1$ T, $\lambda=10$ meV, $D=-255$ meV, $F=1.4 \lambda$ meV, 
$k=290 eV/{\rm nm}^2$ and $d=5\AA$ for three different magnetic field angles. 
}
\end{figure}

In Fig. \ref{fig7} we plot the shift of the resonance frequency as a 
function of $V_{DC}$ for a Ti-H adatom on MgO.  The calculation is carried out 
with the Hamiltonian Eq. (\ref{H}). We take $D=-255$ meV, $F=1.4 \lambda=14.0$
meV, and $B=1$ T.  In order to compute $\delta z$ we take $k=290 eV/{\rm nm}^2$
and a tip-MgO distance of $d=5\AA$. \cite{yang2019} The shift scales inversely
with $d$.  
We consider three orientations for $\vec{B}$, forming angles 
$\theta_B=10^{\circ},45^{\circ},80^{\circ}$.  Expectedly, the resulting 
modulation is linear in $V_{DC}$ and the effect is larger for fields 
off-plane, on account of the fact that $\Delta g_z>\Delta g_x$.   With state 
of the art STM-ESR, the spectral resolution is around 3MHz. Therefore, the 
shift might be observed for large values of $V_{DC}$.   

The exchange with the tip also contributes to the shift in the ESR frequency 
\cite{lado2017,yang2017,yang2019}.  Unlike the g factor modulation mechanism, 
the exchange contribution is tip dependent and decays exponentially with $d$. 
Therefore, for larger tip-surface distance $d$, the $g$ factor modulation 
should dominate.   
Electric shift of the spin resonance was  
observed experimentally in  bulk MgO doped with Cr \cite{royce1963}, by means 
of conventional EPR.

\section{Discussion and conclusions\label{wrap}}

The main idea of this paper is that electric fields can modulate the $g$ 
tensor anisotropy of magnetic adatoms, and that could be used to drive spin 
transitions. Our work comes motivated by recent ESR-STM experiments
\cite{\STMEPR}.   The electric modulation mechanism  discussed here is an 
atomic scale version of the modulation of the $g$ factor of  
electrons\cite{kato2003,malissa2004} and holes\cite{prechtel2015} in 
semiconductor nanostructures,  that lies at the heart of  some well known 
spintronics \cite{datta1990} and is becoming a resource in the manipulation of 
spin qubits\cite{kawakami2014,laucht2015} .  At the atomic scale, the  
modulation proposed here occurs by controlling the weight of two orbital 
states with $\ell_z=\pm 2$, that have a different orbital coupling.   

We now briefly discuss some points of our work that could be improved.
The derivation of the crystal field parameters for Ti-H/MgO  could be 
improved  using a Wannierization \cite{ferron15,lado2017}.   
In addition, we could also improve the model 
Eq. (\ref{H}) by including the effects of hybridization between the $d$ 
orbitals of Ti and the $p$ and $s$ orbitals of oxygen and $s$ orbitals of 
Titanium and Hydrogen, as well as the effect of charge 
fluctuations\cite{Ferron_Lado_prb_2015}. 

Our estimation of the piezoelectric stretching could be improved in 
several ways. First, we are treating the silver substrate and most of the MgO 
as completely rigid. We have verified that keeping the MgO layer completely 
rigid, or letting a few atoms of the MgO layer close to the Ti adatom relax, 
has a minor impact on our estimate of $k$.  We have also treated the potential 
as quasi-static.  Whereas this is probably a good approximation, it the 
Gigahertz frequency might resonate with long wavelengths off-plane MgO phonons 
that would definitely  change the Ti-tip distance, very important for the exchange modulation mechanism\cite{lado2017,yang2019}, and perhaps the Ti-O 
distance as well. 

The $g$ factor observed experimentally for Ti-H/MgO is $g\simeq 1.8$ for 
magnetic fields pointing almost in plane.  From our theory we expect a larger 
value, closer to $g_x=1.97$.  There are two possible reasons for this discrepancy. First,  the 
coupling to silver, ignored both in our DFT calculation and in the model, 
could distort slightly the electronic cloud of Ti-H, which would in turn 
change $D$ and $F$, and thereby $g_x$ and $g_z$. Second, and also related to 
silver, the Kondo coupling to the substrate electrons is expected to 
renormalize $g$, in analogy to the Knight shift\cite{langreth1972theory,
Delgado_Hirjibehedin_sc_2014}. In the case of an isotropic interaction, the 
renormalization of the $g$ factor reads, up to first order in the Kondo 
exchange  between the adatom and the substrate electrons,
 $J_s$:
\begin{equation}
\delta \Delta_z= -\frac{\Delta_z}{2} g_s \rho J_{s}
\end{equation}
where $\rho_s$ is the density of states of substrate at the Fermi energy and 
$g_s$ is the $g$ factor of the substrate electrons. The sign of $J_s$ is 
positive for antiferromagnetic exchange, which is expected in this system, 
since Kondo effect was observed for Cu/MgO \cite{yang2018}.  Therefore, the 
Kondo interaction could reduce the $g$ factor of Ti-H.

We now summarize the main results of this work:
\begin{enumerate}
\item We have shown that an anisotropic  time-dependent modulation of the $g$ 
tensor induces electronic spin transitions,  described by Hamiltonian Eq. 
(\ref{Rabi0}) and characterized by a Rabi coupling given in Eq. (\ref{Rabi1}).
\item We have worked out an analytical theory for the anisotropic $g$ factor 
of the Ti-H $S=1/2$ adatom on MgO and we have benchmarked it against DFT 
calculations. Our theory relates $g$ with crystal field parameters $D$ and 
$F$, as well as Ti spin orbit coupling $\lambda$ (Eq. (\ref{gz}), 
Eq. (\ref{gx}) and Fig. \ref{fig3bc}). 
\item We have computed the modulation of the $g$ tensor anisotropy due to the 
piezoelectric strain of the Ti-H chemisorbed on an oxygen atom on MgO  and we 
have estimated the resulting Rabi coupling (see Fig. \ref{fig4}).  We have 
found that is much smaller than the one observed experimentally, confirming 
the dominance of the exchange modulation mechanism\cite{lado2017,yang2019}.  
However, we have shown that for heavier adatoms with much larger spin orbit 
coupling $\lambda$, this mechanism could be efficient.
\item We have studied to what extent  the crystal field mechanism for ESR-STM, 
proposed by Baumann {\em et al.} to understand the experiments for Fe on 
MgO\cite{Baumann2015}, could be  ascribed to the modulation of the $g$ factor. 
We find that the dominant contributions to the crystal field mechanism come 
from the modulation of the zero field splitting parameters.  
\item  We have also proposed that a DC voltage can shift the $g$ factor, and 
thereby the Zeeman splitting and we have computed the effect for the case of 
		Ti-H (see Eq. (\ref{shift:eq}) and Fig. \ref{fig7}).
This shift provides an additional knob to fine tune the resonance frequency.
\item We have discussed the impact on the hyperfine coupling of the $g$ factor 
anisotropy and its electric modulation (see Eq. (\ref{h1hf})).
\item We have discussed the role of the $g$ factor anisotropy of  Ti-H/MgO to 
	enable the exchange-modulation mechanism for ESR-STM (Eq. (\ref{rabiJ-aniso})). 
In the absence of adatom $g$ factor anisotropy, the exchange-modulation 
ESR-STM mechanism can only work if the tip moment is not aligned with the 
applied field (see Eq. (\ref{anisotip}) ). 

\end{enumerate}

In this work we have focused mostly on Ti-H/MgO and Fe/MgO, although most of 
the ideas can be applied, or extended to the case of other atoms.  Most 
notably, the case of Cu/MgO will be the subject of a future publication. 

{\em Acknowledgments}
We acknowledge Kai Yang,  for fruitful discussions.  J. F.-R. acknowledges 
financial support from FCT  Grants No. P2020-PTDC/FIS-NAN/4662/2014,    UTAP-EXPL/NTec/0046/2017, as well as Generalitat Valenciana funding 
Prometeo2017/139 and MINECO-Spain (Grant No. MAT2016-78625-C2). 
A. F.  acknowledges hospitality from the Department of Physics of Univeridad de 
Alicante. A. F., S. A. R. and S. S. G. acknowledge financial support from
CONICET (PIP11220150100327, PUE22920170100089CO).

\appendix
\section{Calculation of Rabi constant for spin model\label{Rabiapp}}

In this appendix we compute the Rabi coupling defined in eq. (\ref{Rabi0}).   
We consider the general situation for a $S=1/2$:
\begin{equation}
{\cal H}= \vec{b}_0\cdot\vec{S}_1 + \vec{b}_1(t)\cdot \vec{S}
\end{equation}
where 
\begin{equation}
\vec{b}_0=|\vec{b}_0|(\sin \theta_0, 0,\cos\theta_0)
\label{b0}
\end{equation}
and
\begin{equation}
\vec{b}_1(t) =|\vec{b}_1| \cos(2\pi f t) (\sin \theta_1, 0,\cos\theta_1)
\label{b1}
\end{equation}
We shall give explicit expressions for $\vec{b}_0$ and $\vec{b}_1$ below, where we consider independently the exchange modulation and the $g$-factor modulation mechanism.

The eigenstates of $H_=\mu_B\vec{b}_0\cdot\vec{S}$ are, satisfy $H_0|\pm\rangle =\pm \mu_B|\vec{b}_0||\pm\rangle$ and are given by:
\begin{eqnarray}
|+\rangle= \cos\frac{\theta_0}{2} |\uparrow\rangle +\sin \frac{\theta_0}{2} |\downarrow\rangle 
\\
|-\rangle= \sin\frac{\theta_0}{2} |\uparrow\rangle -\cos \frac{\theta_0}{2} |\downarrow\rangle 
\end{eqnarray}

We obtain the matrix element of the spin operators in the basis of eigenstates:
\begin{equation}
\langle +|S_x|-\rangle= \frac{1}{2}\left(\sin^2\frac{\theta_0}{2}-\cos^2\frac{\theta_0}{2}\right)
=-\frac{1}{2}\cos\theta_0
\end{equation}

\begin{equation}
\langle +|S_z|-\rangle= \sin\frac{\theta_0}{2} \cos\frac{\theta_0}{2} = \frac{1}{2} \sin\theta_0
\end{equation}

We can now write the general expression:
\begin{equation}
\Omega= \frac{|\vec{b}_1|}{2} \cos(2\pi f t)
\left( -\sin \theta_1 \cos\theta_0 + \cos\theta_1 \sin \theta_0\right)
\end{equation}
This can be further simplified to:
\begin{equation}
\Omega= \frac{|\vec{b}_1|}{2}\sin(\theta_1-\theta_0)=\frac{1}{2}
\frac{\vec{b}_0\times\vec{b}_1}{|\vec{b}_0|}
\label{RabiGENERAL}
\end{equation}

\subsection{Expression for $\Omega$ for the $g$ factor modulation}
We  now apply eq. (\ref{RabiGENERAL}) for the case of the $g$ factor modulation. 
We now write up 
\begin{equation}
\vec{b}_0= \mu_B \left( g_x B_x, 0,  g_z B_z\right)
\end{equation}
and
\begin{equation}
\vec{b}_1= \mu_B \left(\delta g_x B_x, 0, \delta g_z B_z\right)
\end{equation}

Explicitly, we write:
\begin{equation}
\sin\theta_0=\frac{g_x B_x}{\sqrt{ (g_x B_x)^2 + (g_z B_z)^2}}
\label{sin}
\end{equation}
and
\begin{equation}
\cos\theta_0=\frac{g_z B_z}{\sqrt{ (g_x B_x)^2 + (g_z B_z)^2}}
\label{cos}
\end{equation}

We thus write:
\begin{equation}
\Omega=\frac{\mu_B}{2} \left( -\delta g_x B_x  \cos\theta  + \delta g_z  B_ z \sin \theta\right)
\label{Rabi1A}
\end{equation}

We now use eq. (\ref{sin}) and (\ref{cos}) to express $B_x$ and $B_y$ in terms of $\theta, g_x,g_z,B_x,B_z$ and we obtain:
\begin{equation}
\Omega=\frac{\mu_B}{2}|\vec{b}_0| \cos\theta_0 \sin \theta_0
 \left( -\frac{\delta g_x}{g_x}   + \frac{\delta g_z}{g_z}   \right)
\label{Rabi1B}
\end{equation}
Now we use $\cos\theta_0 \sin \theta_0=\frac{1}{2}\sin2\theta_0$ to obtain:
\begin{equation}
\Omega=\frac{\mu_B}{4}|\vec{b}_0|  \sin 2\theta_0
 \left( \frac{\delta g_z}{g_z}   - \frac{\delta g_x}{g_x}   \right)
\label{Rabi1C}
\end{equation}

We now write the Zeeman splitting 
\begin{equation}
\Delta_Z=\mu_B |\vec{b}_0|
\end{equation}
so that we obtain the expression:
\begin{equation}
\Omega=\frac{\Delta_Z}{4}  \sin 2\theta
 \left( \frac{\delta g_z}{g_z}   - \frac{\delta g_x}{g_x}   \right)
\label{Rabi1D}
\end{equation}

\subsection{Expression for $\Omega$ for the exchange modulation\label{rabiJsec}}
We now consider a Hamiltonian for a $S=1/2$ surface spin where, in addition to the Zeeman interaction, there is a exchange coupling to the magnetic moment of the tip:
\begin{equation}
{\cal V}_{\rm exch} = J(z) \vec{n}_T\cdot\vec{S}
\end{equation}
where
\begin{equation}
\vec{n}_T= \left(\cos(\theta_B+\delta),0,\sin(\theta_B+\delta)\right)
\end{equation}
and $\vec{B}=B_0 \left(\cos(\theta_B),0,\sin(\theta_B)\right)$

We define the spin splitting 
\begin{equation}
\Delta_Z^*\equiv \sqrt{ (g_x \mu_B B_x+ Jn_x)^2 + (g_z\mu_B B_z+J n_z)^2}
\end{equation}
and we express the angles as
\begin{equation}
\sin\theta_0=\frac{g_x B_x+ J n_x }{\Delta_Z^*}
\label{sinJ}
\end{equation}
and
\begin{equation}
\cos\theta_0=\frac{g_z B_z+J n_z }{\Delta_Z^*}
\label{cosJ}
\end{equation}

For the time dependent component, we now {\em ignore} the modulation of the $g$ factors and we only consider the modulation of the exchange, that we write up as $\delta J \cos2\pi f t$. We thus can write
\begin{equation}
\vec{b}_1(t)= \delta J \left( n_x , 0, n_z\right)
\end{equation}
After some algebra we obtain:
\begin{equation}
\Omega_J=\frac{\delta J}{2\Delta_z^*} \left(g_x \mu_B. B_x n_z - g_z \mu_B B_z n_x\right)
\end{equation}

Since we are interested in the role of the $g$ factor anisotropy, we make it explicit and we write:
\begin{eqnarray}
g_x= g +\Delta g_x
\nonumber \\
g_z= g +\Delta g_z
\end{eqnarray}
where $\Delta g_x$ and $\Delta g_z$ are the  static  contributions to the $g$ factor anisotropy.  
We now define
\begin{equation}
{\cal E}_J\equiv \frac{\delta J}{2\Delta_z^*} g\mu_B B
\end{equation}
so that the expression for the Rabi reads:
\begin{eqnarray}
\Omega_J={\cal E}_J
 \left(\cos\theta_B \sin(\theta_B+\delta )-\sin\theta_B \cos(\theta_B+\delta) \right)
+\nonumber
\\
{\cal E}_J
 \left(\frac{\Delta g_x}{g}  \cos\theta_B \sin(\theta_B+\delta )
 -\frac{\Delta g_z}{g}
 \sin\theta_B \cos(\theta_B+\delta)\right)
 \nonumber
\end{eqnarray}
We can now write this up as:
\begin{eqnarray}
\Omega_J={\cal E}_J \sin \delta
+\nonumber
\\
{\cal E}_J
 \left(\frac{\Delta g_x}{g}  \cos\theta_B \sin(\theta_B+\delta )
 -\frac{\Delta g_z}{g}
 \sin\theta_B \cos(\theta_B+\delta)\right)
 \nonumber
\end{eqnarray}
After some algebra we obtain: 
\begin{eqnarray}
\Omega_J={\cal E}_J \sin \delta
+\nonumber
\\
{\cal E}_J \left(\frac{\Delta g_x}{g} -\frac{\Delta g_z}{g}\right)
 \cos\theta_B\sin\theta_B \cos\delta+\nonumber\\
+{\cal E}_J\left(\frac{\Delta g_x}{g} \cos^2\theta_B+\frac{\Delta g_z}{g}\sin^2\theta_B
\right)
\sin\delta
\end{eqnarray}

\section{Estimation of SOC from NIST Database\label{soc:est}}
From the NIST database\cite{nist1999}, we obtain the experimental values for Ti(IV) with an outermost electronic configuradion $d^{1}$.  The lowest energy levels have $L=2$ and $S=1/2$,  with $J=3/2$ and $J=5/2$. Their energy splitting is $\Delta E=47.3$ meV.  We can relate this to the spin orbit coupling using $J=L+S$ and:
\begin{equation}
\lambda\vec{L}\cdot\vec{S}=\frac{\lambda}{2}\left(J(J+1)-L(L+1)-S(S+1)\right)
\end{equation}
From here we obtain:
\begin{equation}
E(J)= \frac{\lambda}{2}\left(J(J+1)-L(L+1)-S(S+1)\right)
\end{equation}
and
\begin{equation}
E(J+1)-E(J)= \frac{\lambda}{2}\left(J+1(J+2-J)\right)=(J+1)\lambda
\end{equation}
Since $J+1=\frac{5}{2}$ we obtain $\lambda_{IV}=\frac{2}{5}\Delta E_{IV}=18.9$ meV.
If we consider Ti(III), we have $J+1=3$ and $\Delta E_{III}=22.9$meV.  This yields
$\lambda_{IV}=\frac{1}{3}\Delta E_{III}=7.63$

It is apparent that the strength of the atomic spin orbit coupling depends on the charge imbalance in the Ti.  One can expect that value of $\lambda$ for  Ti-H on MgO must be in between these two values.

\section{Relation between the $(\ell^{+})^4+(\ell^{-})^4$ and the $\ell_x^4+\ell_y^4$ terms
\label{CFAPP}}
In this appendix we discuss the connection between these two crystal field operators, that we have used for Ti and Fe. The choice is a matter of convenience.   After some algebra, the relation between these two operators is
\begin{equation}
\ell_x^4+\ell_y^4=\frac{(\ell^{+})^4+(\ell^{-})^4}{8}+24 I -\frac{91}{12} \ell_z^2 +\frac{7}{12}\ell_z^4
\end{equation}
Thus, it is apparent that the last two terms can can be reabsorbed as a renormalization of the  $D\ell_z^2$ term, plus a  shift of the $\ell_z=0$ level, that plays a very minor role in the discussion for Ti-H/MgO.

\bibliography{biblio2}{}

\end{document}